\documentclass[pdflatex,sn-mathphys-ay]{sn-jnl}

\usepackage{graphicx}%
\usepackage{multirow}%
\usepackage{amsmath,amssymb,amsfonts}%
\usepackage{amsthm}%
\usepackage{mathrsfs}%
\usepackage[title]{appendix}%
\usepackage{xcolor}%
\usepackage{textcomp}%
\usepackage{manyfoot}%
\usepackage{booktabs}%
\usepackage{algorithm}%
\usepackage{algorithmicx}%
\usepackage{algpseudocode}%
\usepackage{listings}%
\usepackage{graphicx}
\usepackage{subfigure}
\usepackage[table]{xcolor}
\usepackage{tikz}
\usetikzlibrary{arrows.meta,positioning}
\usepackage{xurl}  
\UseRawInputEncoding

\newcommand{\subsubsubsection}[1]{\paragraph{#1}}

\begin{document}

\title[The First Observations of Moonlit Satellites]{The First Observations of Moonlit Satellites}

\author*[1,2]{\fnm{Sarah E.} \sur{Caddy}}\email{sarah.caddy@unimelb.edu.au}

\author[1]{\fnm{Stephen E.} \sur{Catsamas}}

\author[1]{\fnm{Michele} \sur{Trenti}}

\author[2,3]{\fnm{Lee R.} \sur{Spitler}}

\author[4]{\fnm{Patrick} \sur{North}}

\affil[1]{\orgdiv{School of Physics}, \orgname{The University of Melbourne}, \country{Australia}}
                
\affil[2]{\orgdiv{Australian Astronomical Optics}, \orgname{Macquarie University}, \country{Australia}}

\affil[3]{\orgdiv{Astrophysics and Space Technologies Research Centre}, \orgname{Macquarie University}, \country{Australia}}

\affil[4]{\orgname{Ansys}, \country{USA}}

\abstract{
Optical Space Domain Awareness (SDA) operations have traditionally relied on Sunlit passes near the terminator, limiting continuous tracking of low-Earth orbit (LEO) satellites. In this work, we show that by accounting for all four major illumination sources: Sunlight, Earthshine, Moonlight, and Lunar-Earthshine, large LEO satellites like the ISS can be optically monitored for a full 24 hour period. We present what we believe to be the first quantitative observations of the International Space Station (ISS) and Chinese Space Station (CSS) illuminated only by Moonlight and Lunar-Earthshine under night-time conditions. Using a $0.6$\,m telescope, we obtain 147 detections of the ISS with a median brightness of $V = 12.02 \pm 0.17~\mathrm{mag}$, approximately $13.1 \pm 1.3$ magnitudes fainter than in daylight. For interpretation, we extend a satellite brightness model based on reflected Sun light (\texttt{lumos-sat}) to include Moonlight and Lunar-Earthshine, and validate this model both against the new observations and independent Ansys Systems Tool Kit simulations, achieving residuals of $0.03 \pm 0.80~\mathrm{mag}$. The simulations confirm that Lunar-Earthshine dominates the illumination of nadir-facing components, boosting their radiance by at least 100 times relative to Starlight alone. Applying this validated model to existing and proposed large satellites, we show that Moonlight and Lunar-Earthshine will make these future spacecraft well within detection limits of even modest ($\sim 0.6$\,m) optical SDA systems for most of the night, and for up to $\sim 11$ nights per lunar month at local midnight. When Moonlit observations are combined with daytime and twilight observations, this enables near-continuous ($\approx 24\,\mathrm{h}$) optical monitoring of LEO satellites from a single site, significantly increasing SDA tracking capabilities.
}


\maketitle

\section{Introduction}

Optical astronomy has largely been shielded from the full impact of satellite light pollution thanks to favourable orbital geometry. Satellites in low Earth orbit (LEO) that impact optical astronomy observations are illuminated by the Sun in ``terminator illuminated'' conditions. This occurs for an hour or two after Sunset depending on the observers latitude, altitude and time of year. Under these conditions, large artificial satellites can outshine even the brightest stars \citep{nandakumar_high_2023}, but are undetectable for the majority of the time astronomical observations take place as they are shadowed by the Earth. While this is beneficial to Astronomers, it presents a challenge to Space Domain Awareness (SDA) operators who use optical sensors to track a satellites location, and monitor its status in orbit.
\\
\\
As larger satellites become more numerous, and astronomy facilities become more sensitive, we enter a new regime where light from the Moon must also be considered when calculating the brightness of satellites throughout an entire night. Unlike Sunlight, Moonlight has the potential to impact optical observations throughout the entire night. In this work, we present the first detections of satellites illuminated entirely by Moonlight and Lunar-Earthshine (defined in this work as Moonlight scattering off the surface of the Earth) in astronomical night conditions. In this work we will explore the potential for this phenomena to be used in persistent monitoring of satellites for SDA throughout an entire night. We also explore if this will have a negative impact on astronomy observations into the future by quantifying the brightness of objects observed by Moonlight. 

\subsection{Increasing Challenges from the Growth of Satellite Constellations}
The number of satellites being launched into orbit is growing exponentially. Based on current trends and the ambition of private space companies, in the next decade alone it is likely that there will be more satellites launched into orbit than in the entire history of human space exploration to date \citep{borlaff_satellite_2025}. While many of these satellites provide critical infrastructure for people on Earth, they are also creating a growing concern surrounding space sustainability \citep{lawrence_case_2022}. A new phenomena of particular concern are satellite mega-constellations. These are networks of satellites on the scale of hundreds to millions of units which provide services like internet and communications in a distributed network, generally inhabiting orbits in LEO. Examples include SpaceX's Starlink \citep{kandula_simulated_2025} (approx. 42,000 units announced), Eutelsat's One Web \citep{littoriano_modelling_2024} (approx. 7000 units announced), AST SpaceMobile's BlueBird \citep{cole_initial_2025} (approx. 243 units announced) and Amazon's Kupier (approx. 3200 units announced). As satellite numbers increase, so too does the probability that a satellite's path through the sky will intersect with the field of view of optical astronomy images \citep{mallama_satellite_2025, hasan_dark_2023, hainaut_impact_2020, venkatesan_impact_2020} and contaminate spectra, particularly integral field spectrographs like the Multi Unit Spectroscopic Explorer (MUSE) on the Very Large Telescope (VLT) which collect spectra across swathes of sky \citep{streicher_sky_2011}. 
\begin{figure}
    \centering
    \includegraphics[width=1.0\linewidth]{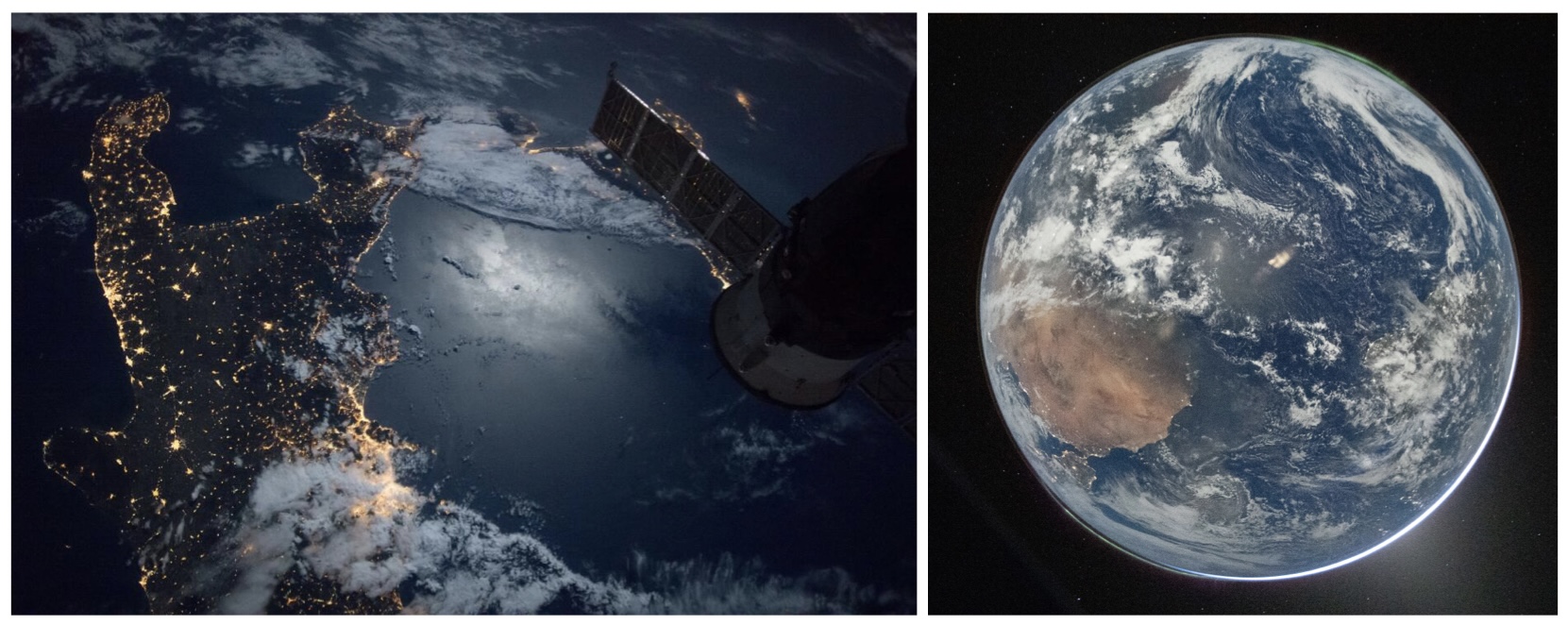}
    \caption{The Moon is an additional source of light which can illuminate satellites at night. Lunar-Earthshine, or Moonlight light scattered off the surface of the Earth, can also illuminate satellites from below. Left: The effect of Lunar-Earthshine is captured in this image taken by the ISS expedition 49 crew on the 17th September 2016. The southern tip of Italy can be seen illuminated by the light of the Moon, as well as a Russian Soyuz spacecraft which can be seen illuminated by direct Moonlight from above, and Lunar-Earthshine from below. Image Credit: NASA. Right: The Earth illuminated by Moonlight, taken by an Artemis II astronaut in the Orion capsule on route to the Moon on the 2nd of April 2026. Image Credit: NASA.}
    \label{fig:iss_moonlight}
\end{figure}
When considering the illumination of satellites for both optical SDA techniques and impact to Astronomy, illumination from Moonlight and Lunar-Earthshine are orders of magnitude fainter than direct Sunlight, and often not considered. The Sun is generally considered to have an apparent magnitude of -26.5, and the full Moon an apparent magnitude of -12.5 \footnote{\url{https://www.atnf.csiro.au/resources/education/senior-astrophysics/photometry/magnitude/}}. 
\begin{equation}
    I_{Sun}/I_{Moon} = 2.5 ^{ m_{Moon} – m_{Sun}} = \sim 400,000
\end{equation}
Thus, satellites in Low Earth Orbit illuminated by the Sun can roughly be considered to be $\sim 400,000\times$ brighter (or 14 magnitudes) than when illuminated by direct Moonlight. For Starlink satellites (V1.5) that have a Sunlit magnitude of $\sim 6$ Vmag at night (depending on spacecraft attitude and phase angle, see \citealt{caddy_daytime_2025}), this equates to a brightness of $\sim 20$ Vmag. 

\subsection{The Rise of Orbital Data Centres}
Satellites illumination by the Moon is faint compared to Solar illumination. However, satellites are getting larger, and astronomical telescopes are becoming more sensitive. In addition to internet and communications constellations, there are now proposals of large Artificial Intelligence (AI) data centre constellations. For example, SpaceX has recently been approved by the U.S. Federal Communications Commission (FCC) to launch 1 million satellites in orbits from 500km to 2000km at 30 degree orbital inclination and sun-synchronous orbits \footnote{\url{https://api-prod.fcc.gov/icfs-attachment/exp/api/v1/46e636811b7afe1011336467624bcbfb}}. Some orbital data centres are proposed to be up to $4km \times 4km$ in size, equating to an angular size on sky of the full Moon at an orbital altitude of 500km \citep{marcy_impact_2026}. 
If we compare these proposed orbital data centres with $16km^2$ reflecting surface area from their solar panels, to the observations of Starlink V1.5 satellites described in \cite{caddy_daytime_2025, fankhauser_satellite_2023} with a solar array surface area of $22m^2$, orbital data centres will reflect $\sim 730,000\times$ more light, or 14.7 magnitudes. This results in an average total integrated brightness across viewing angles of $\sim 5.3$ Vmag. This is comparable to the total integrated brightness of Sunlit Starlink satellites, except with the potential to impact observations all night. Should these targets be within the detectable limit for sensitive survey telescopes like the Vera Rubin Observatory \citep{team_vera_2026}, (as will be explored in this work) the implications could be potentially devastating.
\\
\\
There have been no official reports or studies of observations that characterise satellite brightness when illuminated by the Moon that the authors are aware of at the time of writing. Visual detection of Moonlit Iridium satellites (which, when they were still in orbit produced the famous Iridium flares; \citealt{maley_visual_2003}) have been quietly spoken about in amateur astronomy circles for over a decade \footnote{\url{https://www.cloudynights.com/topic/337979-earth-satellite-illuminated-by-moonlight/}} however no example images or direct measurements have been found. The impact of Moonlight on satellites in orbit can perhaps be very simply illustrated in images taken by NASA astronauts on board the ISS at night. In \autoref{fig:iss_moonlight} (left), a key inspiration for this work, the Moonlight can be seen scattered off the surface of the Earth and potentially illuminating the satellite from below, which we call in this work Lunar-Earthshine. Direct Moonlight can also be seen illuminating the top of the satellite. \autoref{fig:iss_moonlight} (right) is another striking example of the Lunar-Earthshine, and was one of the first images sent back to Earth from the Artemis II mission. Taken by an Artemis crew member from the Orion spacecraft, this image shows the Earth illuminated entirely by the light of the Moon.  
\\
\\
There are no AI data centres in orbit at the time of writing - but the space ecosystem is changing rapidly. In order to explore how large Moonlit satellites might impact astronomy and SDA in the future, we focus on the largest satellite in orbit today, the International Space Station (ISS). In this work we seek to quantify for the first time, the brightness of satellites illuminated by direct light from the full Moon and Lunar-Earthshine from real observations. We present observations of the ISS and the Chinese Space Station (CSS) illuminated by the Moon, and we present a modified version of the {\texttt{lumos-sat}} satellite optical brightness model \citep{fankhauser_satellite_2023} for which we add Moonlight as an illumination source, to compare our observations to model predictions. We then use this model to determine the brightness of both existing and planned large satellite constellations, to understand the potential impact that these satellites may have on future large optical ground based telescope facilities. Finally, we also consider the benefit that these finding may have for SDA operators who wish to track satellites throughout an entire night.

\section{Method}

\begin{figure}
    \centering
    \includegraphics[width=1\linewidth]{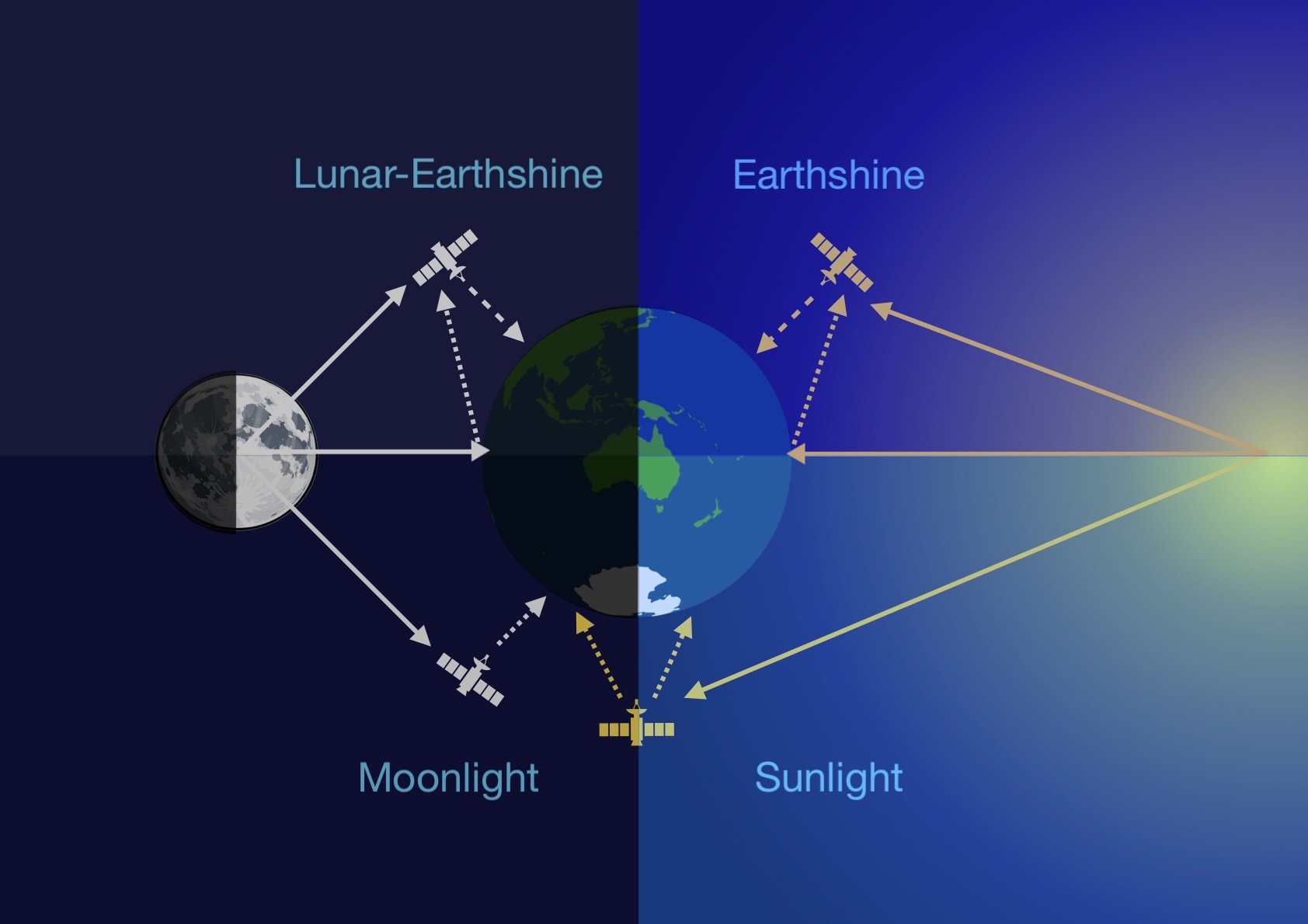}
    \caption{The four dominant sources of illumination for a satellite in LEO can be described as Sunlight: direct light from the Sun which is reflected off the satellite and towards an observer on Earth. Earthshine: light from the Sun which reflected off the surface of the Earth, illuminates the satellite, and then is reflected towards an observer on Earth. Moonlight: Sunlight which illuminates the Moon that is reflected onto a satellite and then directed towards an observer on Earth. We introduce the fourth source as Lunar-Earthshine which is the light from the Moon, reflected onto the surface of the Earth to illuminate the satellite, and then reflected back to an observer on Earth. Which phenomena dominates is dependant on the orientation and location of the satellite, the location of the observer on Earth, and position of the Sun and Moon at the time of observation.}
    \label{fig:moonshine_diagram}
\end{figure}

Throughout this work we will refer to 4 sources of the light which can illuminate satellites in LEO. To reduce confusion, we clearly define each of these sources here. Sunlight is direct light from the Sun scattering off a satellite. Earthshine is the light of the Sun reflecting off the surface of the Earth, and illuminating a satellite. Moonlight is the direct light from the Moon scattering off a satellite. And finally we introduce a new term, Lunar-Earthshine, which is the light of the Moon reflecting off the surface of the Earth and illuminating the satellite. These four terms are illustrated in \autoref{fig:moonshine_diagram}.

\subsection{Description of Observations}\label{obs}

We conduct observations of the ISS (NORAD ID: 25544) and the CSS (NORAD ID: 48274) on nights surrounding, and on full Moon from June 8th to June 12th 2025. We attempt to observe every ISS pass during this period for which the satellite rises above 40 degrees altitude for our observing location, and we are successful in detecting the satellite on all occasions. For this work, we use the 24 inch Planewave research telescope located at Macquarie University Observatory, Sydney, Australia. The instrument consists of the CDK24 f/6.5 optical tube assembly\footnote{\url{https://planewave.com/products/cdk24-ota/}} (OTA), mounted on a L-600 direct-drive telescope mount\footnote{\url{https://planewave.com/products/l-600-telescope-mount/}}, capable of slewing up to 50 degrees per second. The telescope is used both for astronomy outreach and research, and so makes use of a Planewave Series-5 field rotator, IFR90 integrated rotating focuser and Perseus 4-port instrument selector including eyepiece, fibre pick-off which feeds an onsite research spectrograph, planetary imaging camera, and deep sky camera. For this work, we use the deep sky camera, consisting of a ZWO ASI6200MM Pro with field of view (FOV) of $31.0 \times 20.7$ arcmin and a pixel scale of $1''/pix$. This instrument was not able to be equipped with photometric filters at the time of use. This limitation will be discussion in the derivation of the zeropoint.    
\begin{figure}
    \centering
    \includegraphics[width=1\linewidth]{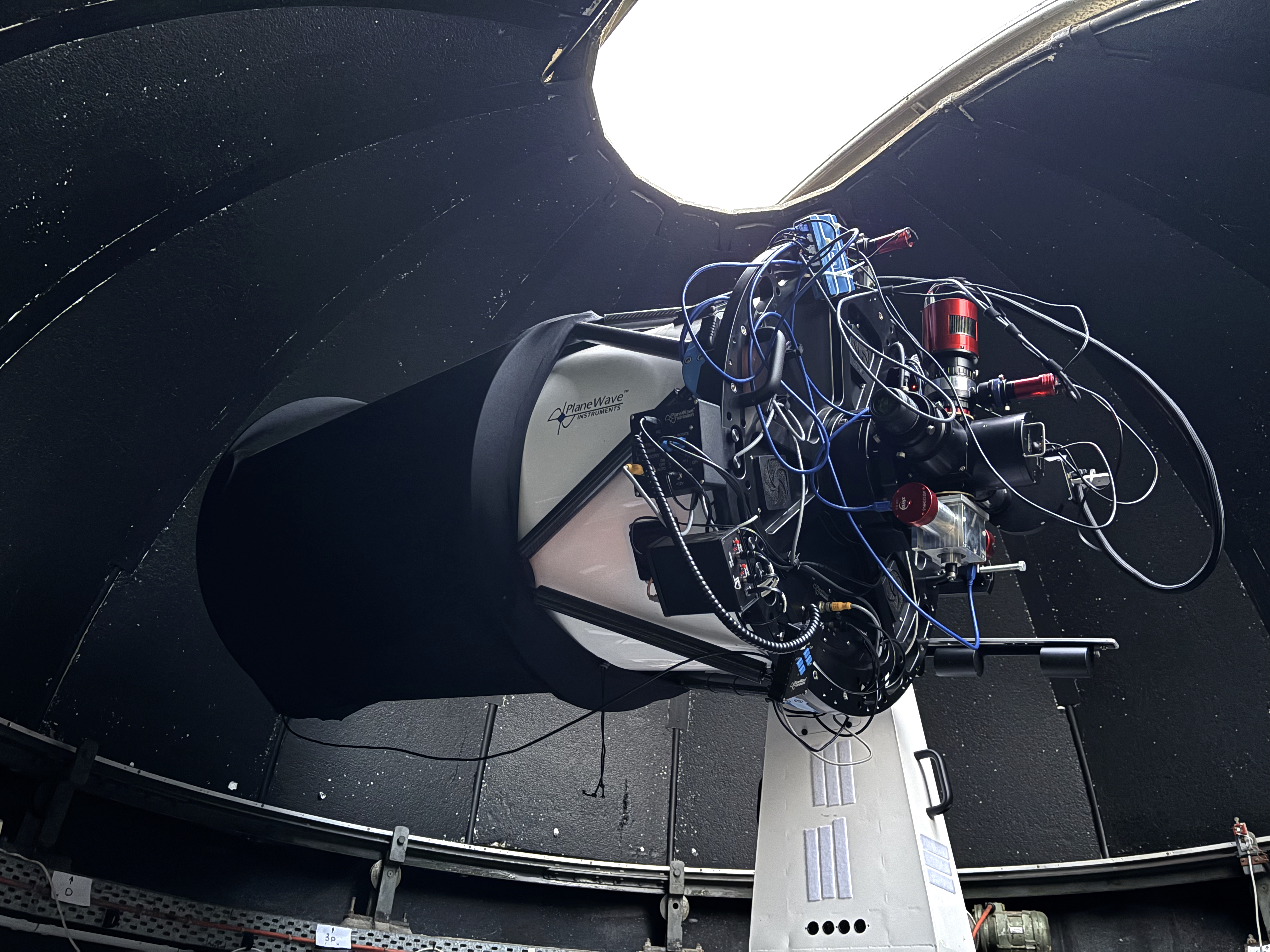}
    \caption{The PlaneWave CDK24 f/6.5 telescope and L600 mount located at Macquarie University Observatory, Sydney, Australia, which is used in this work to observe the International Space Station and Chinese Space Station at night, lit by the Moon.}
    \label{fig:planewave_telescope}
\end{figure}
\\
\\
To actively track the satellites, we use the native PlaneWave Interface 4 software\footnote{\url{https://planewave.wpengine.com/software/}}, and download the Two Line Element (TLE's) manually at the time of tracking via CelesTrak\footnote{\url{https://celestrak.org/}} to ensure they are up to date. An 101 star pointing model has been carried out on the instrument to ensure accurate pointing and tracking to 15.5'' RMS. Google Maps is used to determine the precise longitude and latitude of the telescope, and the observatory altitude is estimated using a topographical map\footnote{\url{https://en-au.topographic-map.com/}}. We note for observers who may wish to reproduce these results, the system is capable of tracking satellites via the popular Software Bisque telescope control suite TheSkyX\footnote{\url{https://www.bisque.com/product/theskyx-pro/}} which is used in our work tracking Starlink satellites during the day with the Huntsman Telescope Pathfinder \citep{caddy_daytime_2025, caddy_optical_2024}. However we find that tracking performance is markedly improved using the native PlaneWave Interface 4 when operating the L600.  
\\
\\
The telescope is located inside a custom, autonomously tracking dome. To prevent vignetting, we do not track below $\sim 15$ degrees. The dome slit is large enough such that the OTA is not obstructed, however an important caveat in this work is that because the dome tracking system is designed for deep sky imaging (not tracking fast moving targets) the dome is limited to discrete tracking pulses. This can result in momentary vignetting when a target is moving quickly, particularly when close to zenith. While it is unlikely that this has impacted the majority of observations, to be conservative, observed magnitudes quoted should be considered as lower limits because of the risk of momentary partial vignetting. Another limitation of these observations is the limited slew rate of the field rotator, particularly towards zenith where the target is moving the fastest. As a result, towards the apex of each pass, the object either moves out of the field of view (so data points are missing from the light curves), or the object is elongated in the image and is not centred. Due to the nature of tracking objects, particularly when using exposures times on the order of seconds, there are times when bright stars can contaminate exposures. As a result of these complications, every image in the observing program has been visually inspected and vetted to ensure its appropriateness for photometric analysis.  
\\
\\
The full detector array of $9576 \times 6388$ pixels is read out. The gain and exposure time is dynamically selected for each frame depending on the brightness of the source and the sky brightness in relation to the target to Moon separation angle. For most exposures, the exposure time is typically set to between 1-10s with gain between 200-300dB to achieve a high enough signal-to-noise (SNR) ratio that the object can be detected by human eye in each frame, while the tracking remains stable enough to reduce smearing of the target across many pixels. The focus is tuned at the start of every run using a star at zenith of comparable magnitude to the satellite target. Altitude and azimuth of the satellite reported in this work is calculated in post processing using the time and date recorded in the FITS header, and the satellite TLE downloaded from CelesTrak. 

\subsection{Data Reduction}

A master flat field is produced by median combing 42 flat frames at an azimuth of 126 degrees taken on the morning of June 11th. Due to the impact of the instrument selector and field rotator, a non-uniform vignetting pattern occurs at the edges of the detector which is azimuth dependent. To avoid the impact of this complication, we clip each field to a central radius of 2500 pixels. This master flat is used to calibrate the science frames and zero point frames. We note that we do not match the gain of the flat field to every science frame (which is variable and is recorded for each frame). Instead we choose a constant gain of 200dB to match the zeropoint frames, and use the manufacturers camera performance metrics\footnote{\url{https://www.zwoastro.com/product/asi6200/}} to convert each frame from ADU to electrons for a given gain level. 
\\
\\
The zeropoint for the system is derived from observations of 321 stars over 21 individual pointings, ranging from an airmass of 1.03 to 2.03. To reduce errors introduced by the lack of a filter, we limit star $B-V$ to between 0.2 and 0.9, assuming the colour of the satellites due to Sunlight and Moonlight reflecting off them to be close to a solar spectrum with a B-V of 0.65 \citep{cox_allens_2002}. To measure the flux of the reference stars we use a median combined image for each pointing of 10-12 frames, each with a gain of 200dB, close the mean gain of the target exposures. The magnitude of stars used range from $\sim 10 - 15$mag V band, many of which would traditionally saturate in photometric catalogues like SkyMapper and SDSS. As a result, we catalogue match stars to the AAVSO Photometric All Sky Survey DR9 using VisieR \citep{henden_vizier_2016}, which is designed to fill this gap. Stars in this catalogue range from V band $10<Vmag<17$ mag for targets fainter than the Tycho2 catalogue (complete to V mag = 11). Calibrated magnitudes can then by found using the following:
\begin{equation}
V = V_{inst} - ZP_V - k_V(X) - C_{V}(B-V)
\end{equation}
\noindent
Where $V_{inst}$ is the instrumental uncalibrated magnitude, $ZP_{V}$ is the photometric zeropoint for V band, $k_V(X)$ is the extinction coefficient for an airmass $X$, and $C_{V}(B-V)$ is a colour correction term derived from catalogue $B$ and $V$ magnitudes. We find a $ZP_{V}$ of $-23.37 \pm 0.11$, a $k_V$ of $0.12\pm0.05$ and a $C_{V}$ of $0.08\pm 0.12$ for the system on the night of the 10th June. We use the same parameters for every observing night. Because the colour term is large in this case, likely due to the lack of a filter, the final photometric error is $\sigma_\text{phot} = 0.17$ given by:
\begin{equation}
    \sigma_{\text{phot}} = \sqrt{
    \sigma_{{ZP_{V}}}^2 +
     \sigma_{k_{V}}^2 +
     \sigma_{C_{V}}^2}
\end{equation}

\subsection{Satellite Brightness Modelling}\label{satmodel}

To extend brightness estimates to fainter targets such as Starlink satellites, brighter targets like AI orbital data centres, and alternative orbits and illumination conditions, we use several methods to validate our observations. First, we present simplified brightness estimates from first principles to explore some fundamental properties of observing large, bright, resolved objects. Secondly, we extend the {\texttt{lumos-sat}} model\footnote{\url{https://github.com/Forrest-Fankhauser/lumos-sat}} described in \cite{fankhauser_satellite_2023} to encompass additional sources of illumination from the Moon. This method also allows us to propagate brightness observations under different illumination conditions for various basic satellite models consisting of 2 planes; a solar panel and chassis component. Finally, we compare our resolved observations of the ISS to a more complex multi-component model using the Ansys Systems Tool Kit (STK) to validate brightness from Lunar-Earthshine and direct Moonlight components.

\subsubsection{Simplified Brightness Model}\label{first_order}

In this work, we describe a convenient, simple approximation for the brightness of a satellite in order to explore some fundamental limitations of observing very faint, resolved satellites.
\\
\\
Depending on the context of the observation, the brightness of a satellite can be expressed in three distinct quantities:
(i) the \emph{total integrated brightness} of a satellite while being tracked;
(ii) the \emph{surface brightness} of a satellite in the regime where
it is spatially resolved; and (iii) the \emph{effective surface brightness}
of a satellite trail produced in a sidereal-tracking astronomical exposure. We define each clearly here.

\subsubsubsection{(1) Total integrated brightness of a tracked satellite}\label{tracked_mag}

When a satellite appears smaller on-sky than the point spread function (PSF) of the telescope system + seeing conditions, the satellite is unresolved. In this case we may integrate all the light from the observation to obtain the total integrated brightness of a satellite. In simple model, a satellite can be approximated as a perfectly diffuse (or Lambertian) reflecting panel. When the satellite is at zenith with respect to the observer, the distance between satellite and observer can be approximated as being equal to the orbital altitude. Under direct full-Moon illumination the satellites brightness is then:
\begin{equation}\label{eq:Fobs}
  F_{\rm obs}
  = \rho \, F_{\rm Moon} \, A_{\rm sat} \,\frac{
    \cos\theta_i \, \cos\theta_e
  }{
    \pi \, d_{\rm sat}^2
  }
\end{equation}
\noindent
Where $\rho$ is the Lambertian reflectance of the satellite surface, $F_{\rm Moon}$ is the full-Moon brightness at the top of the atmosphere (in W\,m$^{-2}$), $A_{\rm sat}$ is the illuminated projected area of the satellite (in m$^2$), $d_{\rm sat}$ is the satellite--observer distance (in m), $\theta_i$ and $\theta_e$ are the incidence and emission angles, respectively, defined with respect to the surface normal.
\\
\\
In a simple case, we select a Moon altitude of
$\alpha_{\rm M}$ above the local horizon, and we assume the satellite panel is facing the observer. Under these assumptions,
we take
\begin{equation}
  \cos\theta_i = \sin \alpha_{\rm M}, \qquad
  \cos\theta_e \simeq 1,
\end{equation}
The magnitude of the satellite when illuminated only by Moonlight can then be expressed V-band magnitudes as:
\begin{equation}
  m_{\rm sat}=
  m_{Moon}
  -
  2.5 \log_{10}
  \left(
    \frac{F_{\rm obs}}{F_{Moon}}
  \right)
  \label{eq:msat}
\end{equation}

\subsubsubsection{(2) Surface brightness of a tracked resolved satellite}\label{surface_bright}

When a telescope tracks a large satellite, and the apparent angular size of the
object is comparable to or larger than the PSF (usually dominated by the atmospheric seeing conditions, and typically between 1-2 arcseconds), a satellite can be treated as a \emph{resolved} extended source. In this regime, the flux is distributed over a large number of pixels on the detector, reducing the per-pixel SNR. In order to calculate detections limits, it is useful to characterise the satellite by a surface brightness
$\mu_{\rm sat}$ (mag/arcsec$^{2}$), to compare directly to instrumental
surface brightness limits, as even relatively bright objects in integrated total magnitude, when spread out over multiple pixels, can be hard to detect if its surface brightness is low.
\\
\\
In this simple approximation, we assume the satellites surface area can be described as a disk of reflecting area $A_{\rm sat}$ and an equivalent diameter $D_{\rm sat}$ such that  $A_{\rm sat} = \frac{\pi}{4} D_{\rm sat}$. At a large distance $d_{\rm sat}$, the angular diameter on the sky in arcseconds is then:
\begin{equation}
  \theta_{\rm sat}
  =
  \frac{D_{\rm sat}}{d_{\rm sat}}
  \times
  \frac{180\times 3600}{\pi}
\end{equation}
And the corresponding angular area in square arcseconds is:
\begin{equation}
  \Omega_{\rm sat}
  =
  \frac{\pi}{4}\,
  \theta_{\rm sat}^2
  \label{eq:Asat_arcsec}
\end{equation}
In this approximation, we consider that the satellite is approximately uniformly illuminated over its surface visible to the observer. As a result, the mean surface brightness $\mu_{\rm sat}$ is related to the integrated apparent magnitude $m_{\rm sat}$ and $\Omega_{\rm sat}$ by:
\begin{equation}
  \mu_{\rm sat}
  =
  m_{\rm sat}
  +
  2.5 \log_{10}
  \left(
    \Omega_{\rm sat}
  \right)
  \label{eq:mu_sat_resolved}
\end{equation}
This defines the \emph{surface brightness} of a tracked, resolved satellite.
\\
\\
An interesting and important result that follows, is that for satellites that are resolved, their surface brightness is independent of their surface area and orbital altitude. Combining Eqs.~\eqref{eq:Fobs}–\eqref{eq:mu_sat_resolved}, and noting that the
angular area of the satellite on the sky scales as:
\begin{equation}
  \Omega_{\rm sat} \propto \frac{A_{\rm sat}}{d_{\rm sat}^2}
\end{equation}
We obtain:
\begin{equation}
  \mu_{\rm sat}
  =
  m_{\rm Moon}
  - 2.5\log_{10}\rho
  - 2.5\log_{10}(\cos\theta_i\cos\theta_e)
  + \mathrm{const.}
\end{equation}
where the explicit dependence on $A_{\rm sat}$ and $d_{\rm sat}$ cancels.
Thus, for a toy model of a uniformly illuminated panel with diffuse scattering properties that is spatially resolved, the surface brightness is independent of the satellite’s physical size and
orbital altitude. The brightness of the target in simple terms, then only depends on the brightness of the illumination source, the surface properties of the reflecting surface, and the illumination geometry. This phenomena in astronomy is known as surface brightness invariance.

\subsubsubsection{(3) Effective surface brightness of a satellite streak}

For sidereal-tracking astronomical exposures, the telescope does not follow
the satellite. Instead, the satellite moves through the field of view and leaves a
trail or streak on the detector. In this case, we consider an effective surface brightness $\nu_{sat}$ $(mag/arcsec^2)$ of the streak, proportionate to the surface-brightness of a satellite streaking through an image, observed over some effective exposure time or pixel crossing time. This is a useful term to calculate, in order to determine if a satellite is visible to within some surface brightness limit of an astronomical survey telescope. 
\\
\\
Following \cite{borlaff_satellite_2025}, the angular width of the satellite trail can be approximated as:
\begin{equation}
  \theta_{\rm sat}^2
  =
  \left(
    \frac{D_{\rm sat}^2 + D_{\mu}^2}{d_{\rm sat}^2}
  \right)
  +
  \sigma^2
  \label{eq:theta_trail}
\end{equation}
Where $D_{\rm sat}$ is the equivalent satellite diameter, $D_{\mu}$ is the telescope mirror diameter, $d_{\rm sat}$ is the satellite to observer distance, and $\sigma$ is the angular resolution of the telescope dictated by environmental effects from seeing conditions, wavelength observed and telescope diameter. The effective time that the satellite takes to cross the detector is:
\begin{equation}
  t_{\rm eff}
  =
  \frac{\theta_{\rm sat}}{\omega_{\rm sat}}
  \label{eq:teff}
\end{equation}
Where $\omega_{\rm sat}$ is the apparent angular velocity of the satellite that an observer sees from the vantage point of the surface of the Earth, as dictated by its orbital altitude and viewing geometry.
\\
\\
As the satellite is moving across the field of view, its flux is effectively averaged over the exposure time $t_{\rm exp}$ and its trail area $A_{\rm trail}$. The resulting surface brightness $\Sigma_{\rm sat}$ (in $W/m^{2}/Hz/arcsec^{2}$) can be written as
\begin{equation}
  \Sigma_{\rm sat}
  =
  \frac{F_{\rm sat}}{A_{\rm trail}}
  \times
  \frac{t_{\rm eff}}{t_{\rm exp}}
  \label{eq:Sigma_sat}
\end{equation}
\noindent
Finally, the effective surface-brightness magnitude of the satellite trail
$\nu_{\rm sat}$ $(mag/arcsec^{-2})$ is:
\begin{equation}
  \nu_{\rm sat}
  =
  -2.5\log_{10}(\Sigma_{\rm sat})
  - 56.1
  \label{eq:mu_sat_streak}
\end{equation}
\newline
Where the constant $56.1$ converts $\Sigma_{\rm sat}$ expressed in
W\,m$^{-2}$\,Hz$^{-1}$\,arcsec$^{-2}$ to AB magnitudes per square arcsecond,
following the convention in \cite{borlaff_satellite_2025} such that:
\begin{equation}
m_{AB} = -2.5 log_{10} (F_{Jy}) + 8.90 \approx -2.5 log_{10}\Big(\frac{F_{Jy}}{3630.78Jy}\Big)
\end{equation}
The AB magnitude system is a direct conversion between flux and magnitude, with that flux being constant in all photometric filters. Note that we will use AB magnitudes throughout this work unless directly specified. 

\subsubsection{Lumos-sat, Two Component Model}

A more complex brightness model is needed to capture satellite geometries, scattering geometries (not just Lambertian as we assumed in the simple model in \autoref{first_order}), and reflectance properties of different materials, which is the dominant factor needed to be considered when regarding larger satellites that may be bright enough to be tracked by Moonlight. {\texttt{lumos-sat}} is the only open source model that the authors are aware of at the time of writing that includes the contribution of Earthshine to the brightness of a satellite, which in \cite{caddy_daytime_2025} we demonstrate to be critically important to the predicted brightness. {\texttt{lumos-sat}} however does not currently include contributions due to the direct light of the Moon, nor the contribution of Lunar-Earthshine scattering off the surface of the Earth. The model does not account for atmospheric Rayleigh scattering, however a modified lunar constant derived from bottom of the atmosphere solar spectra is considered to account for general atmospheric absorption. We add these components to the model as described below. 
\\
\\
In {\texttt{lumos-sat}} we construct satellites as a simplified geometric model consisting of two surfaces, one for the total area of the solar panels, and one for total surface area of the chassis (or in the case of orbital data centres the radiators). Each surface of the satellite model has a Bidirectional Reflectance Distribution Function (BRDF) \citep{nicodemus_directional_1965, greynolds_general_2015} associated with it to describe the angular distribution of light scattered from the surface, unique to the properties of the material. The BRDF takes the general form:
\begin{equation}
 BRDF = f_{r}(\hat{w}_{i}, \hat{w}_{o}) \equiv \frac{1}{cos\,\theta_{e}\,cos\,\theta_{i}\,L_{i}}\frac{\partial{L_{o}}}{\partial{\hat{w}}_{i}}	  
\end{equation}
\noindent
Where $L_{o}$ is the outgoing radiance, and $L_{i}$ is the incoming radiance. $\theta_{i}$ is defined as the angle between the surface normal and the vector to the source $\hat{w}_{i}$, and $\theta_{e}$ is the angle from the surface normal to the vector of the observer $\hat{w}_{o}$. Simply, the BRDF is the function that describes the ratio of spatially distributed radiance to the incident irradiance of a surface. The simplest model being the lambertian BRDF defined as:
\begin{equation}
    BRDF = \frac{\rho}{\pi}
\end{equation}
Where $\rho$ is the albedo of the surface. This is used to describe a diffuse surface that scatters light equally in all directions with no specular reflections. This is a good approximation for surfaces such as the bottom face of BlueBird and BlueWalker satellites\footnote{\url{https://spacenews.com/operational-ast-spacemobile-satellites-could-proceed-without-prototype/}} that are a matte white colour and diffuse surface \citep{cole_initial_2025}. 
\\
\\
For highly specular mirror surfaces like the reflective sails of Earendil-1 (see \autoref{sat_desc}), we approximate the BRDF as a specular reflector, which takes the form of a dirac delta function: 
\begin{equation}
    BRDF = \frac{\rho\, \delta \,(\hat{w}_{i}-\hat{w}_{r})}{cos\,\theta_{i}} 
\end{equation}
In this work we also consider a BRDF for the surface of the Earth to describe Earthshine and Lunar-Earthshine, and the surfaces of the satellites modelled. Firstly, we use a Phong BRDF model for the Earth's surface as described in \cite{fankhauser_satellite_2023} which consists of a composition of a Lambertian component scattered diffusely in all directions ($\frac{\rho_{d}}{\pi}$) and a specular component. The total BRDF is thus given as: 
\begin{equation}
    BRDF = \frac{\rho_{d}}{\pi} + \rho_{s} \frac{n + 2}{2 \pi} ({\hat{w}_{r}} \, \cdot \, {\hat{w}_{o}})^{n} 
\end{equation}
\noindent
Where $\rho_{s}$ the magnitude of the specular component, the $\rho_{d}$ parameter controls of the magnitude of the diffuse component, and n is the angular width of the specular peak. ${\hat{w}_{r}}$ and ${\hat{w}_{o}}$ are the reflected and outgoing unit vectors respectively. In this work we use the Phong BRDF parameters for the land given and tested in \cite{fankhauser_satellite_2023} and \cite{caddy_daytime_2025} of $\rho_{d} = 0.53$,  $\rho_{s} = 0.28$ and $n = 7.31$. Future works may improve on this approximation by considered variable Phong BRDF parameters, based on the surface properties of the Earth below the satellite, such as large bodies of water.
\subsection{Spectral Modelling of Moonlight}
\begin{figure}
    \centering
    \includegraphics[width=1\linewidth]{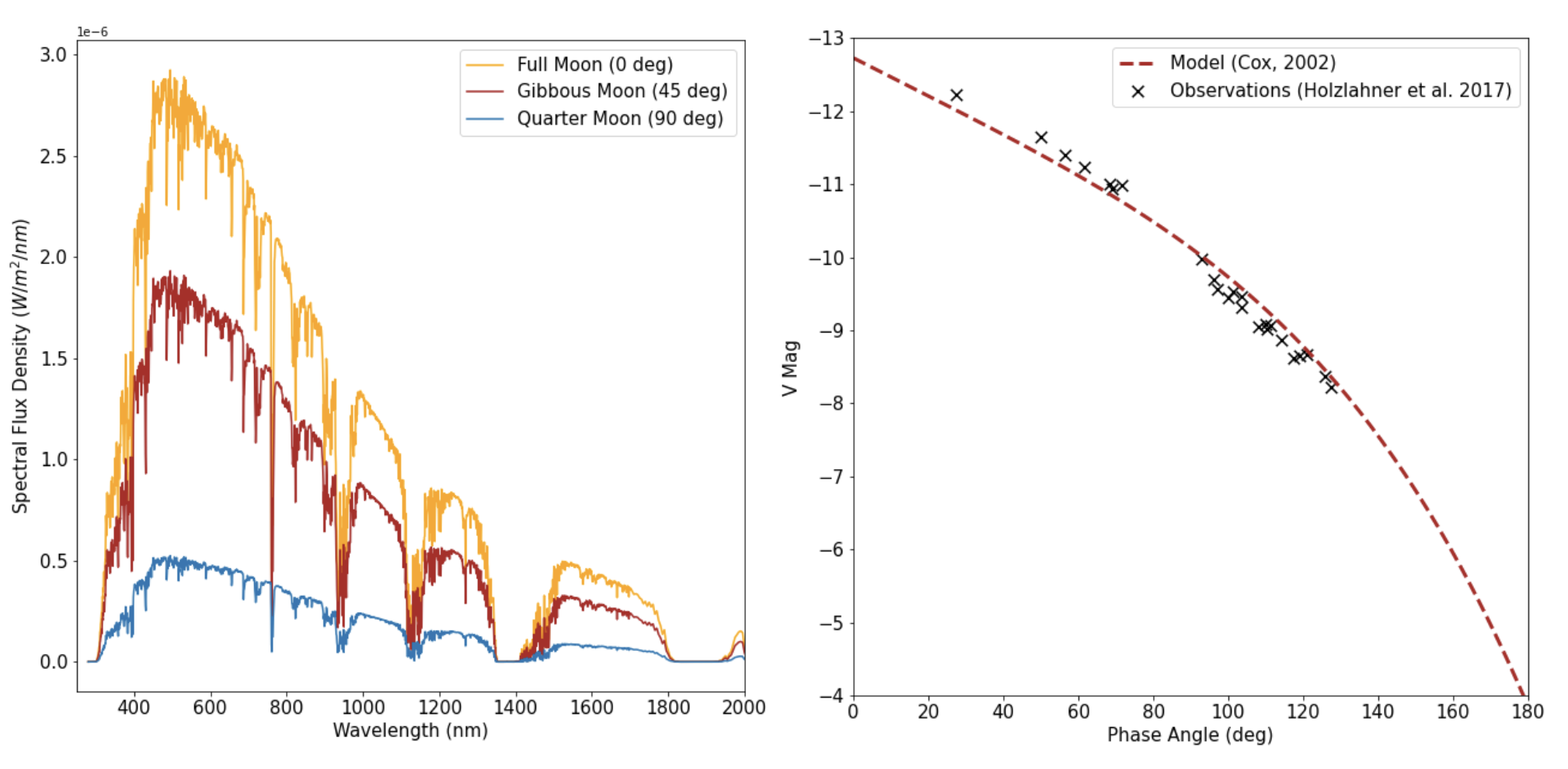}
    \caption{Left: A model for the spectral energy distribution of Moonlight at the bottom of the atmosphere created by scaling an ASTM G-173-03 solar spectrum to the V band magnitude of the Moon for different lunar phase angles. Right: A comparison of the \cite{cox_allens_2002} lunar brightness model used in this work to real Moon observations from \cite{holzlahner_towards_2017}}
    \label{fig:moonshine_model}
\end{figure}

To add Moonlight as an external illumination source in {\texttt{lumos-sat}}, we define a model for the brightness of the Moon as a function of lunar phase angle. This is given in \cite{cox_allens_2002} as:
\begin{equation}\label{moon_bright}
    m = -12.73 + 0.026 |\alpha| + 4\times10^{-9} \alpha^{4}
\end{equation}
\\
\noindent
Where $\alpha$ is the lunar phase angle at the time of observation in degrees, and -12.73 is the maximum observed V band ABmag of the Moon as observed from the surface of the Earth. We compare this model to V band data of Moon's magnitude as a function of phase angle presented in \cite{holzlahner_towards_2017} and shown in \autoref{fig:moonshine_model}. There is good agreement except for smaller phase angles, which may be impacted in part by the opposition effect described in \cite{krisciunas_model_1991} where by at lunar phase angles $< 7$ degrees, the brightness may deviate from this relation, however we assume this widely accepted standard model defined by \cite{cox_allens_2002} as a reasonable first order estimate. We use this computed magnitude for a given lunar phase angle to scale the Lunar spectral energy distribution (SED) in order to compute the total integrate flux in $W/m^2$ incident on the surface of the Earth, or a satellite, for any defined bandpass. We approximate the Lunar SED as a standard solar spectrum (ASTM G-173-03) at the bottom of the atmosphere, to account for atmospheric absorption\footnote{\url{https://www.nrel.gov/grid/solar-resource/spectra-am1.5}} similarly to \cite{caddy_daytime_2025}. The input for the brightness of an illumination source in {\texttt{lumos-sat}} is then the total integrated flux in $W/m^2$ in the desired bandpass, or for imagery without filters, integrated across the telescope throughput.  
\\
\\
\noindent
Using the input brightness and sky position of the Moon, contribution to the total flux at the observer from Moonlight and Lunar-Earthshine are calculated similarly to solar contributions. Therefor, for a lunar phase angle of $\alpha$ and solar phase angle of $\beta$, the total flux at observer is given by:
\begin{equation}
    I_{total}(\beta, \alpha) =  I_{Earthshine}(\beta) + I_{Sunlight}(\beta) + I_{LunarEarthshine}(\alpha) + I_{Moonlight}(\alpha)
\end{equation}

\subsection{Satellites Considered}\label{sat_desc}

Modelled satellites in this work are several constellations with large numbers of total units planned, that also have publicly available information about the size, shape and material properties of the satellite units. These include Starlink satellite models from V1.5 through to V3, Bluebird FM1 satellites, and the first iteration of Reflect Orbital satellites Earendil-1. We also consider large AI data centre platforms with similar publicly available information, including the first iteration of the SpaceX AI data centre satellites called Starmind, and the proposed Starcloud AI data centre. In addition we consider a model of the ISS to compare to observations from public descriptions of the surface area and material properties of the modules on the space stations. The satellites considered in this work are presented in \autoref{tab:satellite_surface_area}. All satellites with solar panel components are assumed to be Sun pointing for nominal operations at all times in the model. Chassis are taken to be nadir pointing.
{
\renewcommand{\arraystretch}{1.4} 
\begin{table}[t]
  \centering
  \begin{tabular}{lccc}
    \hline
    \textbf{Satellite Version} & \textbf{Total Surface Area (m$^2$)} & \textbf{Number Launched} & \textbf{Number Planned} \\
    \hline
    Starlink V1.5      & 25.65 &     2971    &    2971    \\
    Starlink V2.0      & 116.03  &     7838    &   7838     \\
    Starlink V3.0      & 420 &    58     &   15,000     \\
    Bluebird FM1       & 253  &    10     &    40-60    \\
    Earendil-1         & 324  &    0     &    50,000    \\
    Starcloud          & 16,000,000  &    0     &    1    \\
    Starmind           & 1400  &    0     &   1,000,000     \\
    ISS                & 4250.25  &   1      &    1    \\
    \hline
  \end{tabular}
  \caption{The total surface area (solar array + chassis + external radiators if present), number of in-orbit satellite units and the current estimate for the total number of units planned, for each satellite modelled in this work. References for these values and the date of the estimate are described in \autoref{sat_desc}.}
  \label{tab:satellite_surface_area}
\end{table}
}
\\
\\
\noindent
Starlink is currently the largest constellation in orbit. It totals 10,413 operational units as of June 1st 2026\footnote{\url{https://www.space.com/spacex-starlink-satellites.html}}, spanning V1 to V2 generations \footnote{\url{https://planet4589.org/space/con/star/stats.html}}, in predominately 550km altitude orbits with various orbital inclinations. We do not consider the smaller population of V1 Starlinks in this work. The dimensions and configuration of the Starlink satellites from V1.5 and V2 are taken from \cite{caddy_daytime_2025} and consist of a solar panel that is Sun tracking, and a perpendicular chassis facing the Earth in nominal operations. Starlink V3 are approx $4\times$ the surface area of the previous V2 generation \footnote{\url{https://www.nextbigfuture.com/2024/03/spacex-starship-launched-starlink-gen-3-unfolded-nearly-as-wide-as-the-space-station.html}} and at the time of writing are beginning test deployment from Starship in mid July 2026 \footnote{\url{https://spaceflightnow.com/2026/07/16/live-coverage-spacex-to-deploy-first-starlink-v3-satellites-on-suborbital-starship-super-heavy-flight/}}, initially inhabiting an orbital inclination of $33^\circ$ to $53^\circ$, and 525km to 535km orbital altitude \footnote{\url{https://space.skyrocket.de/doc_sdat/starlink-v2-0-ss.htm}}. For the Starlink satellite models, we use the Binomial BRDF derived from laboratory measurements of V1.5 Starlink satellites given in \cite{fankhauser_satellite_2023} for the chassis and solar panels. 
\\
\\
The Bluebird FM1 satellite is a technology pathfinder for AST SpaceMobile’s new generation of communications satellites operating in LEO. They are smaller in total surface area than the Starlink V3, and significantly exceed brightness limits put in place by the International Astronomical Union (IAU) \citep{mallama_satellite_2025, cole_initial_2025}, and are likely to be of concern due to secondary illuminations sources like Moonlight. Dimensions are taken from modelling and observations of \cite{cole_initial_2025}, and we use a diffuse lambertian scattering BRDF to approximate the Bluebird. We note that the current Bluebird model, any deployed solar panels (while unconfirmed and not visible in current model diagrams) will have a negligible projected area to the observer compared to the chassis and will not be well aligned for specular reflections to reach the observer. Therefore we do not consider a solar panel component for this satellite. The Bluebird FM1 will inhabit $53^\circ$ inclination orbits, at 520km, with some satellites in the block 2 configuration will have the ability to raise their orbit from, 530km to 700km altitude, although it is unclear how many satellites will inhabit this higher orbital altitude\footnote{\url{https://apps.fcc.gov
/els/GetAtt.html?id=375091&x=}}.
\\
\\
Earendil-1 is, by design, a deployed, thin film, planar mirror surface for reflecting Sunlight onto the Earth to ``extend daylight" in targeted areas. While the company Reflect Orbital has ambitions to launch a constellation of up to 50,000 units by 2035 \footnote{\url{https://www.reflectorbital.com/}}, currently the FCC has only approved the flight of 1 unit for tests purposes. Details of Earendil-1 dimensions and material properties can be found in the original FCC filing for the orbital debris assessment report \footnote{\url{https://docs.fcc.gov/public/attachments/DA-26-706A1.pdf}}. We only consider the mirror surface components for Earendil-1 BRDF and approximate it as a planar mylar mirror surface of $96\%$ reflectance, and described by a dirac delta function with $0.53^\circ$ beam divergence\footnote{\url{https://orbitalsolar.ai/}}. The satellite will inhabit an orbital inclination of $88^\circ$ with an orbital altitude of 625km.
\\
\\
Starcloud is an envisioned future AI orbital data centre with the largest proposed solar and radiator surface areas of all of the satellites modelled in this work. The total surface area is estimated at 4km x 4km in order to produce 5 gigawatts of power \citep{feilden_why_2024}. Little is known about the proposed orbit, other than targeting a low-Earth, dawn-dusk, sun-synchronous orbit to maximise solar power generation. With lack of information about the satellite, we will approximate the solar panel, chassis and radiator BRDF's from \cite{fankhauser_satellite_2023}.
\\
\\
Starmind in SpaceX's answer to building orbital data centres, and will consist of a distributed system designed to perform artificial intelligence workloads across 1,000,000 individual satellite units lifted into orbit using Starship. Starmind is planned to operate at orbital altitudes ranging from 500 km to 2000 km in both $30^\circ$ and sun-synchronous orbit inclinations inhabiting orbital shells spanning up to 50 km each\footnote{\url{https://docs.fcc.gov/public/attachments/DA-26-113A1.pdf}}. Unlike Starcloud, despite their smaller surface area \footnote{\url{https://www.spacex.com/spacexai/starmind\#power-generation}} these satellites may present a greater concern to astronomers because they will be Sunlit at higher altitudes for longer durations into the night, and visible all night with the potential for Moonlight illumination at all altitudes, in the $30^\circ$ inclination shells.  
\\
\\
Finally, in the absence of BRDF information for the ISS, we also adopt the solar panel and chassis BRDF from \cite{fankhauser_satellite_2023}. This is a reasonable first estimate given that ISS is made of a similar materials (predominantly solar panels and aluminium). In order to accurately estimate the total surface area of the ISS solar array, radiators, truss, modules and visiting vehicles\footnote{\url{https://www.nasa.gov/international-space-station/space-station-visiting-vehicles/}} at the time of observation, we consider every component individually, and refer to the reference guide to the International Space Station \citep{kitmacher_reference_2015} for detailed descriptions of their diameter and length. The sum of these components (which can been seen by an observer on Earth) are taken as the estimate for the surface area of both chassis and solar panel components. The tables in \autoref{secA1} present a detailed description of all of the components of the ISS considered in this work and their quoted dimensions. The total surface area of the solar arrays (including both main, Zarya and visiting vehicles) is $2672.36 m^2$ and the total surface area of the chassis (including radiators) visible to an observer on the surface of the Earth is $1577.89m^2$.

\section{Results}


\subsection{Observational Results}

\begin{figure}[t!!!]
    \centering
    \includegraphics[width=1\linewidth]{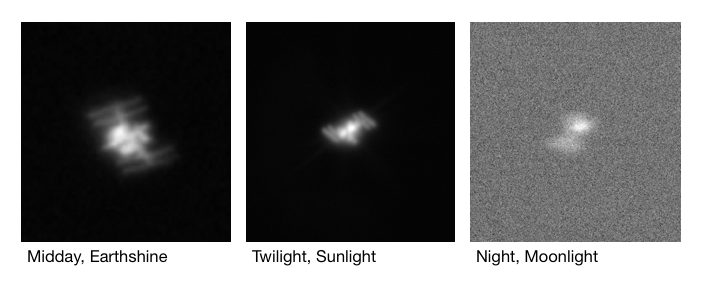}
    \caption{A comparison of observations of the International Space Station at optical wavelengths lit by different light sources. Left is an observation take at midday by the Huntsman Telescope and lit by Earthshine. In the centre is an observation taken during twilight conditions from this work and lit by the Sun, and on the right is an observation taken at night from this work lit entirely by the Moon.}
    \label{fig:iss_clips}
\end{figure}

We present the first observations of the ISS publicly reported at the time of writing, lit entirely by the light of the Moon. In \autoref{fig:iss_clips} we present a visual qualitative comparison of the difference in brightness between midday observations of the ISS taken by The Huntsman Telescope from \cite{caddy_daytime_2025} (left) lit by Earthshine, twilight observations undertaken in this work (centre) lit by direct Sunlight (and potentially also Earthshine - see \cite{fankhauser_satellite_2023} for this analysis) and night observations (right) lit by the light of the full Moon. Even when lit by Moonlight, one can still make out the characteristic shape of solar panels, chassis and bright reflection from the radiators. 
\\
\\
When comparing the magnitude of observed satellites that are actively tracked, we define the magnitude of a target to be the total integrated flux in an aperture encompassing the entire object, in order to compare to {\texttt{lumos-sat}} calculations. Unless otherwise stated, magnitudes are reported in V band ABmags. We do not consider any range corrections for these magnitudes, as our work is focused on the observed brightness of satellites from the perspective of an observer on the ground, not the intrinsic brightness often used for satellite classification and comparison. In this initial work we compare observations of the ISS under twilight, and full night conditions on days spanning Moon phases of $91 - 100\%$ illumination. Future work will expand these illumination conditions. We also compare these observations to images taken in the day given in \cite{caddy_daytime_2025}. 
\\
\\
In total, we compare 147 observations of the ISS lit by the Moon, 52 in twilight conditions, and 363 in daylight conditions. We note that these observations span different phase angles and satellite attitudes and so are not direct comparisons, but allow us to make order of magnitude comparisons over multiple passes.  
\begin{table}[t]
  \centering
  \renewcommand{\arraystretch}{1.4}
  \begin{tabular}{lccc}
    \hline
    \textbf{Time of Day} & \textbf{Illumination Source} & \textbf{Median Magnitude} & \textbf{Standard Deviation} \\
    \hline
    \rowcolor{gray!35}
    Nighttime & Moonlight + Lunar-Earthshine & $12.02\pm0.17$ & $1.00\pm0.17$ \\
    \rowcolor{gray!20}
    Twilight & Sunlight & $6.41\pm0.17$ & $1.93\pm0.17$ \\
    \rowcolor{gray!5}
    Daytime & Sunlight + Earthshine & $-1.10\pm0.05$ & $0.84\pm0.05$ \\
    \hline
  \end{tabular}
  \caption{A summary table of the brightness of the ISS under different illumination conditions. Mean magnitudes and standard deviation are presented with photometric errors. Daytime observations are taken from \cite{caddy_daytime_2025} and have a lower photometric error.}
  \label{tab:satellite_brightness_tab}
\end{table}
\\
\\
In \autoref{tab:satellite_brightness_tab} we report the median brightness of the ISS for daylight, twilight, and full night observations. Observations of the ISS taken at full night with no direct Sunlight incident on the satellite have a median magnitude of $V=12.02\pm0.17$ with $\sigma = 1.00\pm 0.17$ Vmag where $\pm 0.17$ is the derived photometric zeropoint error and $\sigma$ is the standard deviation of the observations. For observations taken during twilight conditions, we find a bimodal distribution which is caused by some of the observations being taken in direct Sunlight, and some taken in the transition period between direct Sunlight illumination and diffuse scattered Sunlight and Earthshine scattered in the atmosphere. 
\\
\\
In twilight conditions, we find a median magnitude of $V=6.41\pm0.17$ and $\sigma = 1.93\pm0.17$ Vmag with $\sigma$ being the largest for these observations due to the diverse range of illumination conditions on twilight. We also compare these observation to daytime magnitudes from \cite{caddy_optical_2024} which report $V=-1.10\pm0.05$ and $\sigma = 0.84\pm0.05$ Vmag which originated from direct Sunlight scattering off the satellite, as well as Earthshine. Note, photometric uncertainties in observed magnitudes are assumed and not reported for the rest of the work. We note that the observation taken during the day used the Huntsman Telescope Pathfinder were undertaken at the same observing site to this study, so the seeing conditions (accounting for difference between night and day observation), observatory altitude, and the surface features of the Earth which contribute to Lunar-Earthshine and Earthshine will be similar \citep{caddy_optical_2024, caddy_daytime_2025} however will change with differing local cloud cover. 
\begin{figure}[t!!]
    \centering
    \includegraphics[width=0.8\linewidth]{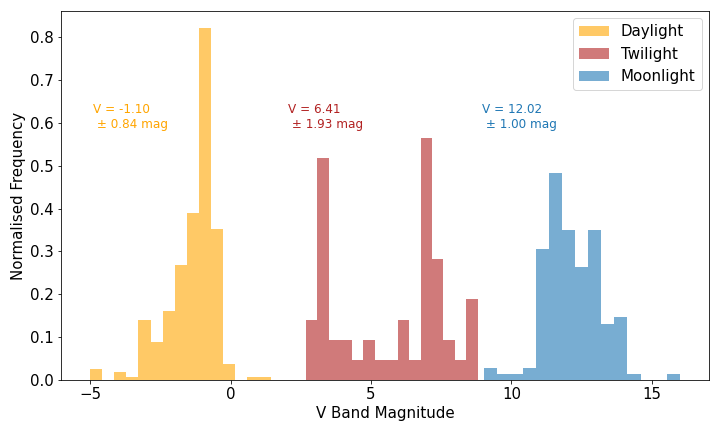}
    \caption{The distribution of observed brightness of the ISS under different illumination conditions. We include 147 observations of the ISS taken at night, 52 observations are included at twilight under diverse illumination conditions resulting in a bimodial distribution and 363 daytime observations of the ISS taken with the Huntsman Telescope Pathfinder \citep{ caddy_daytime_2025}.}
    \label{fig:iss_moonlight_brightness_dist}
\end{figure}
\\
\\
From these results we note that the ISS is found to be $13.12\pm1.30$ magnitudes fainter at night than when observed during the day, which is within the order of magnitude estimate of 14 magnitudes derived from the observe magnitude difference of the Sun and Moon brightness. Variances are likely to originate from the differences in the geometries of the passes observed at night and during the day, resulting in different illumination angles. The corresponding surface brightness of the ISS is then $\mu = 20.60 mags/arcsec^2$.
\begin{table}[b]
  \centering
    \setlength{\tabcolsep}{2pt}
    \begin{tabular}{lcccccc}
    \hline
    \textbf{Time (local)} & \textbf{Target} & \textbf{Moon Phase (\%)} & \textbf{Sun Alt ($deg$)} & \textbf{Illum} & \textbf{Max (Vmag)} & \textbf{Med (Vmag)}\\
    \hline
    24-01-24 10:20 & ISS & 98\% & 50 & Day & -4.73 & -0.83 \\
    08-06-25 20:33 & ISS & 91\% & -44 & Night &  9.00 & 11.91 \\
    09-06-25 02:00 & CSS & 99\% & -59 & Night & 12.94 & 13.42 \\ 
    10-06-25 03:00 & ISS & 99\% & -48 & Night & 10.78 & 11.44  \\
    10-06-25 19:40 & ISS & 99\% & -32 & Night & 11.30 & 12.42 \\
    11-06-25 18:50 & ISS & 100\% & -24 & Twilight & 2.70 & 6.41 \\
    12-06-25 03:00 & ISS & 99\% & -48 & Night &  11.20 & 12.65 \\
    
    \hline
  \end{tabular}
  \caption{The reported max and median Vband magnitude of the ISS and CSS from observations under different illumination conditions and passes.}
  \label{tab:brightness_all}
\end{table}
\noindent
All ISS observations are reported in \autoref{tab:brightness_all}. In addition to the ISS, an observation of the Chinese Space Station is captured at a Moon illumination of $99\%$ and Sun altitude of -59 degrees at 2:00am. We find this object to be fainter at a peak of 12.94 mag and a median of 13.42 mag, likely due to a difference in ratio of highly reflective components (like radiators) to the total solar panel area. Due to lack of public information about the CSS and the limited observations conducted in this campaign we do not model this target with {\texttt{lumos-sat}}, but include brightness results here for completeness. 

\subsection{Comparison to Lumos-sat Model}

In order to predict what satellites can be observed under Moonlit conditions, we validate our changes to the {\texttt{lumos-sat}} model by comparing them with our observations. We consider all 147 observations of the ISS taken at night, and simulate the same illumination conditions in Lumos-sat using the satellite TLE to derive the satellite location in the sky, {\texttt{Astropy}} to determine the position of the Sun and Moon at the time of observation, and \autoref{moon_bright} to determine the brightness of the Moon at the lunar phase for the date of observation. 
\\
\\
Of all the observations, we find the residuals of the observed and modelled brightness to be $0.03\pm0.80$ mag. It is unsurprising that, similarly to the results of \cite{caddy_daytime_2025} the brightness of the satellite predicted in the model across all observations is slightly brighter than that of the observations because as mentioned, we do not take into account any effects of Rayleigh scattering in the atmosphere in this simple model, which is left for future work. 
\begin{figure}[t!!]
    \centering
    \includegraphics[width=1.0\textwidth]{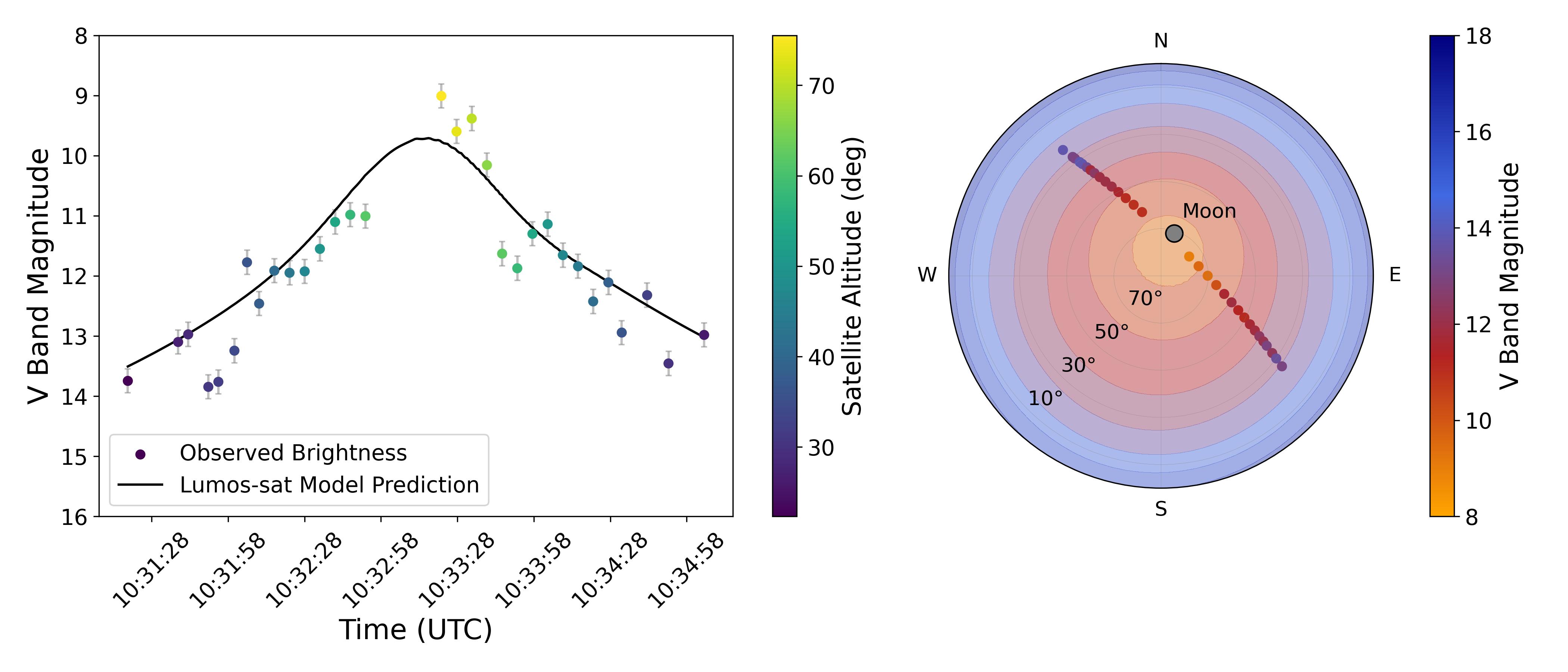}
    \caption{The predict magnitude of the ISS across the entire sky for an observation on the 8th of June. Left: the observed and predicted magnitude as a function of time over the duration of the pass, with colour bar indicating the satellite altitude. Right: an all sky comparison of the predicted and observed magnitude of the ISS. Zenith is the centre of the plot, and the horizon is the edge. The Moon location is indicated with a grey circle. The impact of Lunar-Earthshine can be seen as the source illuminating the satellite from below for an observer on the ground. Observed magnitudes are indicated as a scatter plot, demonstrating agreement with the model.}
    \label{fig:model_comparison}
\end{figure}
\\
\\
In \autoref{fig:model_comparison} left, we compare the ISS pass on the 8th of June to the model prediction. The model prediction follows that shape of the observed light curve closely, with observations at the peak of the pass slightly higher than the model predicts. In \autoref{fig:model_comparison} right we predict the brightness of the satellite across the entire sky with the centre of the graph being Zenith, and the location of the Moon as a grey circle. The observed brightness of the satellite is shown as scattered points on the same colour scale. Again, we see that the satellite observed brightness closely follows the model, with the brightest point being directly below the location of the Moon in the sky - this is expected behaviour for geometry where illumination is driven by Lunar-Earthshine. We can visualise this effect by looking at images taken from the ISS of the Earth at night such as \autoref{fig:iss_moonlight}, which show the bright Lunar-Earthshine as a diffuse glowing spot on the Earth below the Moon, illuminating the ISS from below, and direct Moonlight illuminating the space station from above.
\begin{figure}[t]
    \centering
    \includegraphics[width=1\linewidth]{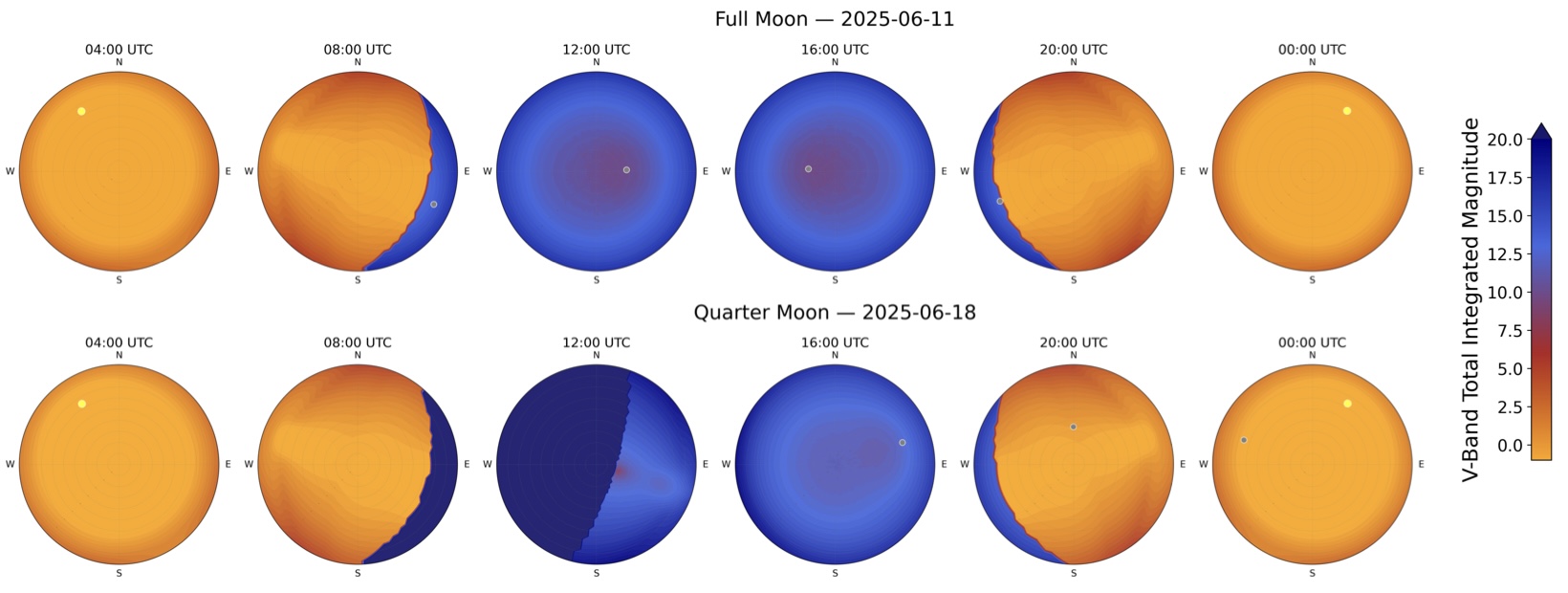}
    \caption{Simulated ISS illumination over a full day–night cycle using the {\texttt{lumos-sat}} model for 11 June 2025 (full Moon) and 18 June 2025 (last quarter). White and grey circles show the Sun and Moon positions respectively, and the colour scale gives the total integrated brightness of the ISS if observed at that point in the sky.}
    \label{fig:all_day}
\end{figure} 
\\
\\
\noindent
There are however, different situations in which direct Moonlight may be visible scattering off the satellite which is observable from Earth. To understand this, we explore the illumination of the ISS using the {\texttt{lumos-sat}} model throughout the entire day-night cycle. This is shown in \autoref{fig:all_day} for the 11th of June 2025 during full Moon conditions, and the 18th of June 2025 during last quarter conditions. In this figure, the location of the Sun is indicated as a white circle, and the location of the Moon as a grey circle. The colour bar indicated the total integrated brightness of the satellite if it were in that location in the sky at the time of observing. Considering full Moon conditions, there are no predicted geometries in which direct Moonlight dominates the illumination conditions. From 04:00 UTC, stepping forward in time by 4 hour time steps, we see the Sun set. At 08:00 UTC the satellite is still illuminated by the Sun during traditional terminator illuminated conditions. Past the terminator, satellites are still illuminated by the Moon. In 12:00 - 16:00 UTC, Lunar-Earthshine is in the predicted dominant component and is brightest directly below the location of the Moon on the surface of the Earth. At 20:00 - 00:00 UTC we see the same conditions reversed in the morning. For case on the 18th of June during quarter Moon, we see that once past the terminator at 08:00 UTC, satellites are not illuminated by any light source. at 12:00 UTC, the Moon has not risen, but we see a bright spectacular reflection, likely from the back of the Sun tracking Solar panels, on the Moon illumination terminator. This progresses to similar illumination conditions as full Moon from 16:00 UTC onwards to morning, albeit with a fainter predicted total magnitude due to the reduction in brightness of the Moon compared to full Moon conditions. 
\\
\\
These models illustrate that the environment in which a satellite is illuminated in LEO is complex, and by considered source of illumination such as Lunar-Earthshine and Moonlight, satellites can be observed in night conditions when they would otherwise be considered too faint for optical sensors. 

\subsection{The Role of Lunar-Earthshine}

The results of \cite{caddy_daytime_2025} suggest that satellites observed during the day are illuminated by Earthshine. This is due to the angle made between the ground based observer, target satellite and source of illumination - in the daytime case, the Sun. It would follow that with similar geometry of a full Moon at night, an illumination source for satellites with a nadir pointing chassis and Sun pointing solar panels is Lunar-Earthshine. 
\\
\\
{\texttt{lumos-sat}} produces total integrated brightness estimates that are in agreement with our observations, but is currently limited to two simple surfaces of different scattering properties in idealised conditions: the solar panels which are Sun tracking, and the chassis which is nadir pointing. The ISS however is a more complex 3D object than a Starlink satellite for example, for which {\texttt{lumos-sat}} was designed to simulate. In order to compare our resolved imagery of the ISS illuminated by Lunar-Earthshine and Moonlight with a model, we simulate one of the observed ISS passes in the Ansys System Tool Kit (STK) \footnote{\url{https://ansys.synopsys.com/products/missions/ansys-stk}}. 
\\
\begin{figure}[t]
    \centering
    \includegraphics[width=0.9\linewidth]{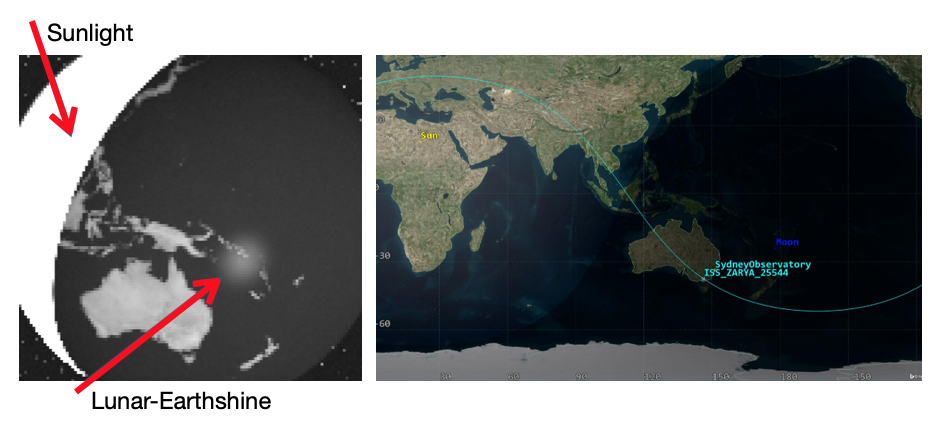}
    \caption{Left: The illumination conditions on the surface of the Earth close to full Moon on the 9th of June 2025 simulated in Ansys STK as seen by a geostationary satellite. The contribution of Lunar-Earthshine can be seen as a specular reflection on the surface of the Earth below the Moon. Right: The orbital geometry of the ISS pass modelled on the 9th of June above Sydney, Australia, with Sun and Moon positions indicated on the map.}
    \label{fig:ansys_setup}
\end{figure}
\\
\noindent
STK is a physics-based software environment which is used to model more complex systems like satellites within realistic illumination conditions. Lunar-Earthshine was not a component of the model natively available, so we approached the Ansys team to make custom modifications to include it for this work. A Lunar-Earthshine model is taken to be $1.0\times 10^{-9} W/cm^2/sr/\mu m$ as measured by a geostationary satellite looking at the Earth at the time if the ISS observation. This is used to create a uniform radiance map across the Earth, simulating Moonlight and Lunar-Earthshine illuminating the ISS. \autoref{fig:ansys_setup} right shows the simulation of the ISS pass considered, and the illumination conditions on the surface of the Earth at the time of observation for an observer in Sydney, Australia. For this simulation, we replicate the peak of the June 9th pass at 10:33 UTC. The figure also shows the location of the Sun and Moon on the surface of the Earth when the observation is made. \autoref{fig:ansys_setup} left shows the illumination conditions simulated, illustrating the specular Lunar-Earthshine spot directly below the Moon (similar to what is simulated in {\texttt{lumos-sat}} in \autoref{fig:model_comparison} left), and the terminator line where Sunlight is directly striking the Earth. 
\begin{figure}[t]
    \centering
    \includegraphics[width=1\linewidth]{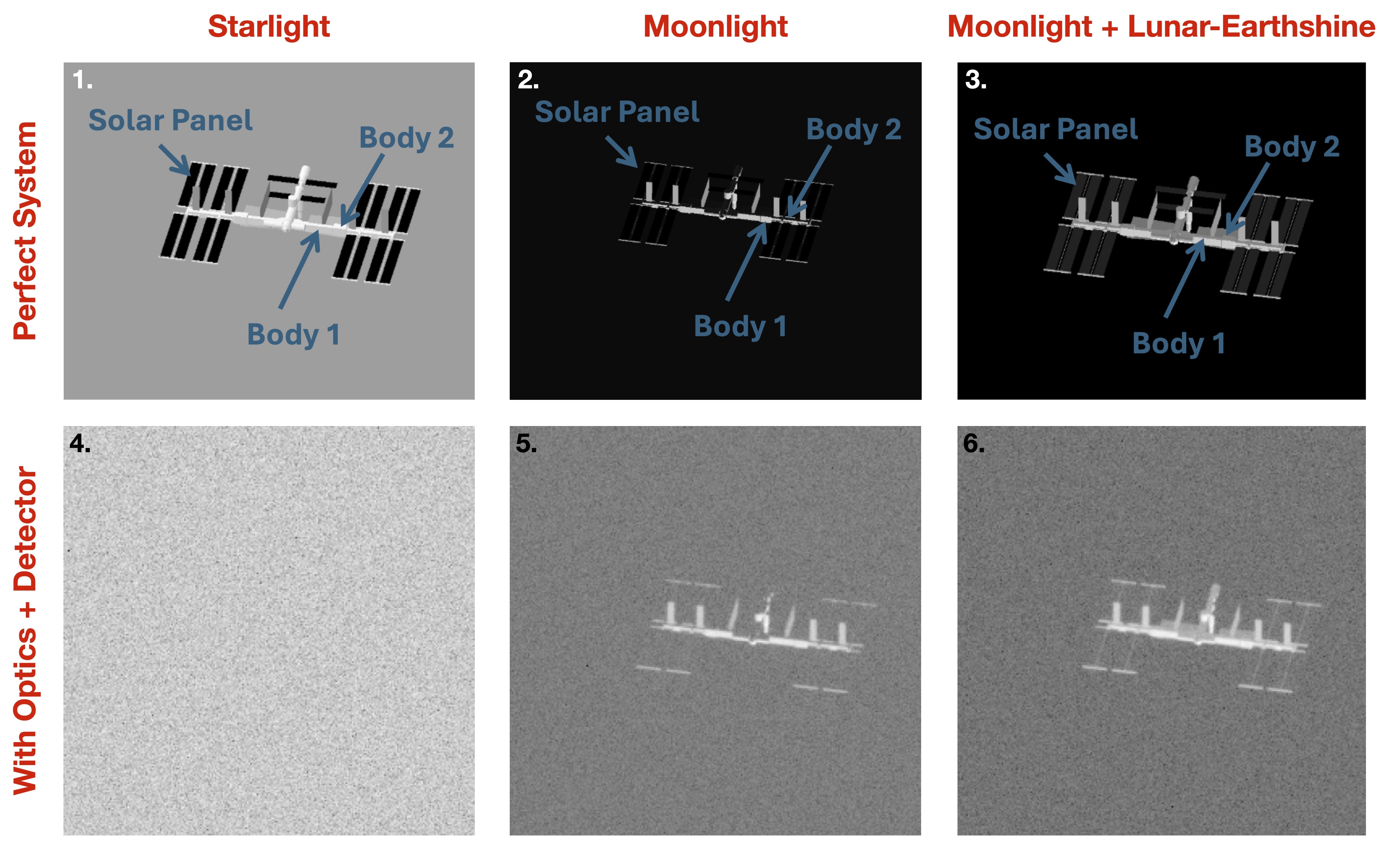}
    \caption{Simulated resolved ISS observations using Ansys STK. Panels 1, 2, and 3 show what the simulated image looks like for the pass on the 9th of June at 10:33 UTC 2025 which was observed in this work. These images assume a perfect entrance aperture radiance and no sensor effects applied, and illumination by starlight, moonlight and Lunar-Earthshine respective. Panels 4, 5, 6 apply the sensor characteristics and impacts of the optical system of the 24" telescope and detector used in this work for the same 3 illumination conditions, in a 30s tracked exposure.}
    \label{fig:ansys_resolved}
\end{figure}
\\
\\
\begin{table}[b]
  \centering
  \setlength{\tabcolsep}{6pt} 
  \setlength{\extrarowheight}{2pt}
  \begin{tabular}{lccc}
    \hline
    \textbf{Value} & \textbf{Starlight Only} & \textbf{Moonlight} & \textbf{Moonlight + Lunar-Earthshine}\\
    \hline
    Solar Panel ($W/m^2/sr$) & $1.48\times 10^{-12}$ &  $1.48\times 10^{-12}$ & $1.38\times 10^{-10}$ \\
    Body 1 ($W/m^2/sr$)      & $1.63\times 10^{-11}$ & $1.35\times 10^{-8}$  & $1.54\times 10^{-8}$ \\
    Body 2 ($W/m^2/sr$)     & $2.36\times 10^{-11}$ & $2.36\times 10^{-11}$ & $4.82\times 10^{-9}$ \\ 
    SNR & 0.575 &  307 & 467 \\
    Irradiance ($W/cm^2$) & $2.87\times10^{-20}$ & $8.45\times10^{-18}$  & $1.31\times10^{-17}$ \\
    Visual Magnitude & 19.63 & 13.46 & 12.98 \\ 
    \hline
  \end{tabular}
  
  \caption{STK ISS radiance and detector measurements for the solar panels, body 1 and body 2 test locations for a pass observed at Sydney on the 9th of June 2025 at 10:33 UTC.}
  \label{tab:iss_bright_rads}
\end{table}
\\
\noindent
The Ansys ISS model includes solar panels, radiators, and docked modules at the time of observation, and assumed Sun tracking solar panels. \autoref{fig:ansys_resolved} shows the results of the simulation. Panel 1 shows a resolved image of the ISS illuminated only by Starlight and with perfect entrance aperture radiance and no sensor effects applied. Panel 2 shows the ISS illuminated by direct Moonlight illumination only at the time of observation and Panel 3 illustrates the same conditions, but with Lunar-Earthshine included in the simulation. Panels 4, 5 and 6 illustrate the same conditions, but apply the sensor characteristics of the telescope system used in this work and described in \autoref{obs}, with a 30s tracked integration between 0.3-1.05 microns. The radiance values of components of the observed satellite (the solar panels, and body 1 and body 2) are given in \autoref{tab:iss_bright_rads}.
\\
\\
We find that no direct Moonlight is incident on the solar panels (the largest reflecting surface area) or body 2 (nadir facing component of the radiators) when Lunar-Earthshine is not considered. For body 1 components, Moonlight increases the brightness compared to starlight only by $\sim830\times$. The solar panels are lit entirely by Lunar-earthshine, and are found to be  $\sim93\times$ brighter than Starlight alone, and body 2 is found to be $\sim240\times$ brighter than Starlight alone and is illuminated only by Lunar-Earthshine. For body 1 which is illuminated by direct Moonlight, Lunar-Earthshine increases the brightness by $\sim 14\%$. This clearly illustrates that Lunar-Earthshine is a non-negligible component of illumination of satellites in Moonlit conditions, and in fact dominates the brightness of nadir facing components. 
\\
\\



\noindent
Applying the sensor characteristics of the telescope used in this work, we find that Moonlight increases the simulated SNR of the ISS from 0.58 (not detected) to 307, and further to and SNR of 467 when Lunar-Earthshine is included. The simulated STK visual magnitude (found using Vega as a reference irradiance of $2.04\times10^{-12}W/cm^2$, calibrated for the atmosphere) is found to be 12.98, matching our median observed value within error margins of $12.02\pm1.00$. The results of this analysis further emphasises the importance of considering Lunar-Earthshine illumination, in particular of nadir facing components of a satellite in LEO, when determining the detection probability during the night.  
\\
\\
Finally, comparing the simulated ISS images to the best resolved image captured in our work shown in \autoref{fig:iss_clips}, the importance of exceptional tracking stability and accuracy, as well as good seeing, is clear. These qualities prevent signal from being averaged or smeared over too many pixels, reducing the SNR of the detection against the bright sky background in full Moon conditions. It is clear from the results of this work, there is much improvement to be made on the tracking accuracy of our system, which will be explored in relation to future work. 

\section{Discussion}\label{disc}

\subsection{Practicality of Moonlit Observations for SDA}

\begin{figure}[t]
    \centering
    \includegraphics[width=0.9\linewidth]{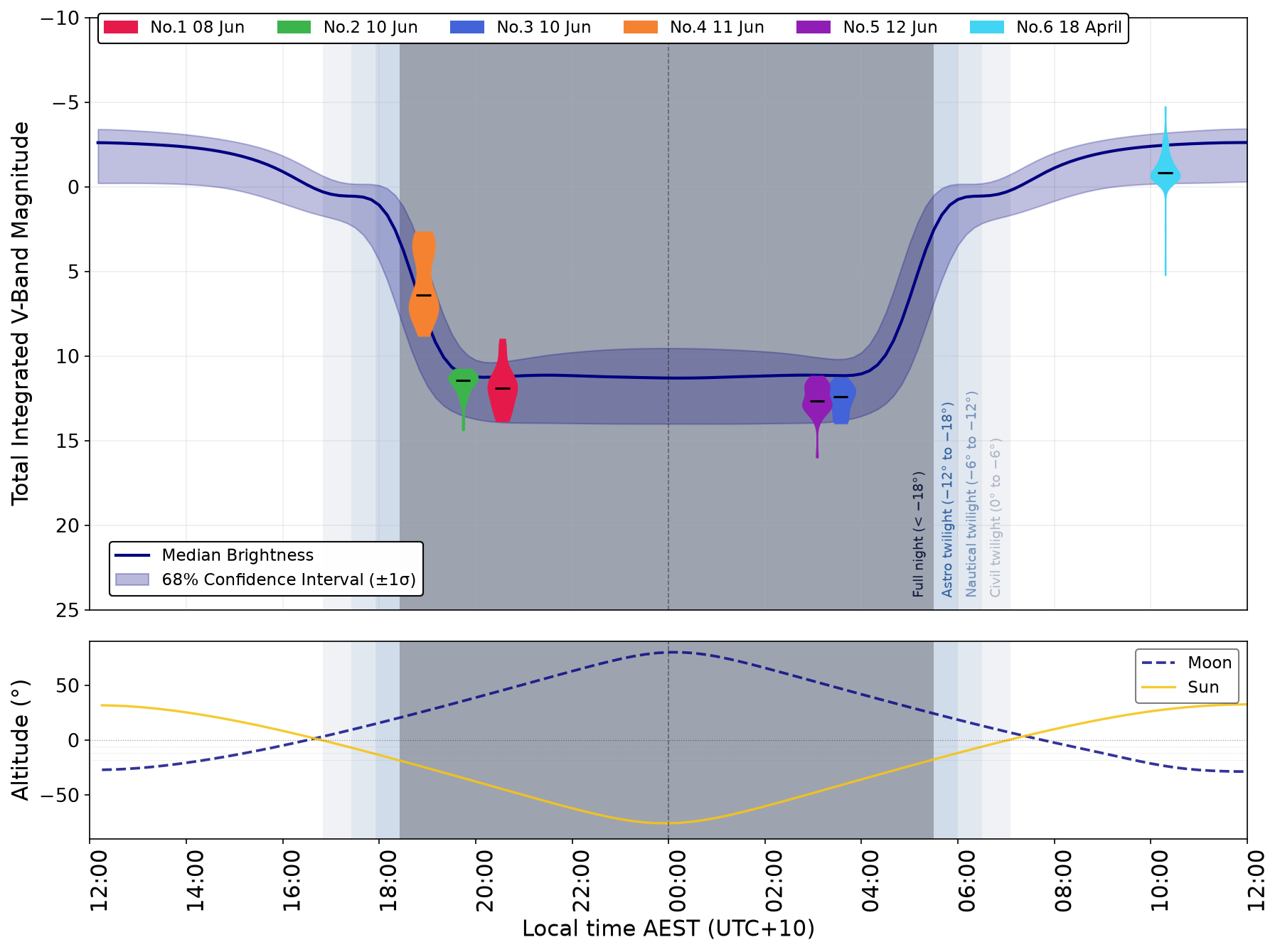}
    \caption{The total integrated brightness of the ISS predicted by the {\texttt{lumos-sat}} model over a 24 hour period during full Moon conditions on the 11th of June 2025. A $1\sigma$ confidence interval is also plotted to show the spread in satellite brightness across the sky for that moment in time. Observational data is plotted as violin plots from this work and daytime measurements from \cite{caddy_daytime_2025} to show the spread of the observed data. A small time offset is applied to observations on the 10th (No.3) and 12th (No.5) of June to make the distributions more easily visible. All median observed brightnesses of the ISS fall within the predicted $1\sigma$ confidence interval. The relative altitudes of the Sun and Moon as shown in the bottom figure for reference.}
    \label{fig:all_day_iss}
\end{figure}

Traditional optical SDA observations taken from the ground usually consist of observing a target when it is still lit by the Sun, but the observatory on the Earth is in shadow. This work has demonstrated for the first time that satellites can be observed when lit by Moonlight, and this has implications for the productivity of ground based optical SDA observatories. To explore this, we use the {\texttt{lumos-sat}} model that has been extended and validated in this work, to calculate the total integrated brightness of the ISS throughout an entire night around full Moon, and compare this to our observations shown in \autoref{fig:all_day_iss}. The brightness of the ISS is taken to be the median of a uniformly distributed number of sampling points in the sky above the observer, at the location of Macquarie Observatory on the night of the full Moon on the 11th of June 2025. The median brightness is then calculated in 10 min intervals throughout the 24 hour period, considering contributions from Sunlight, Earthshine, Moonlight and Lunar-Earthshine. A $1\sigma$ confidence internal is also plotted to illustrate the spread in the brightness across the hemisphere above the observer at any one point in time. For each set of observations considered in this work (one ISS pass made up of multiple individual exposures), we plot a violin plot to show the distribution of brightnesses measured for that pass. 
\\
\\
\noindent
The medians of all observations considered in this work agree well with the model and are found to be within the $1\sigma$ confidence internal for all of the night time, twilight and daytime observations. The ISS is found to be brightest during the day contributions of both Earthshine and Sunlight, with brightness of $-2.66^{-3.44}_{-0.27} \text{ mag}$. The brightness decreases during twilight periods due to the reduction in the contribution of Earthshine (shown in \citealt{caddy_daytime_2025}) before dropping to a Moonlit value throughout the night of $11.25^{9.52}_{13.97} \text{ mag}$. Based on the results of our observation campaign and comparison to the {\texttt{lumos-sat}} model, we find that the ISS is detectable by our modest telescope system throughout the entire night, twilight and daytime during full Moon conditions. As a result in full moon conditions we can observe the ISS for effectively $100\%$ of the 24 hour period, as opposed to twilight + daytime conditions where $\sim 54\%$ of the time is used for observations as shown in \cite{caddy_daytime_2025}, and tradition twilight only conditions when $\sim12.5\%$ of the time utilised for observations (considering an observer in Sydney at the same June period as these observations were taken).
\\
\\
While our observational data points were limited during this test run due to the limited number and times of ISS passes above our location at full Moon (every pass was attempted and was successful in detection), based on the promising results of this preliminary work we will aim to expand this research over multiple epochs to fill in the full 24 hour period, as well as expanding the work to other lunar phase angles in future.    

\subsection{Detection Thresholds Throughout the Month}

\begin{figure}[t]
    \centering
    \includegraphics[width=0.9\linewidth]{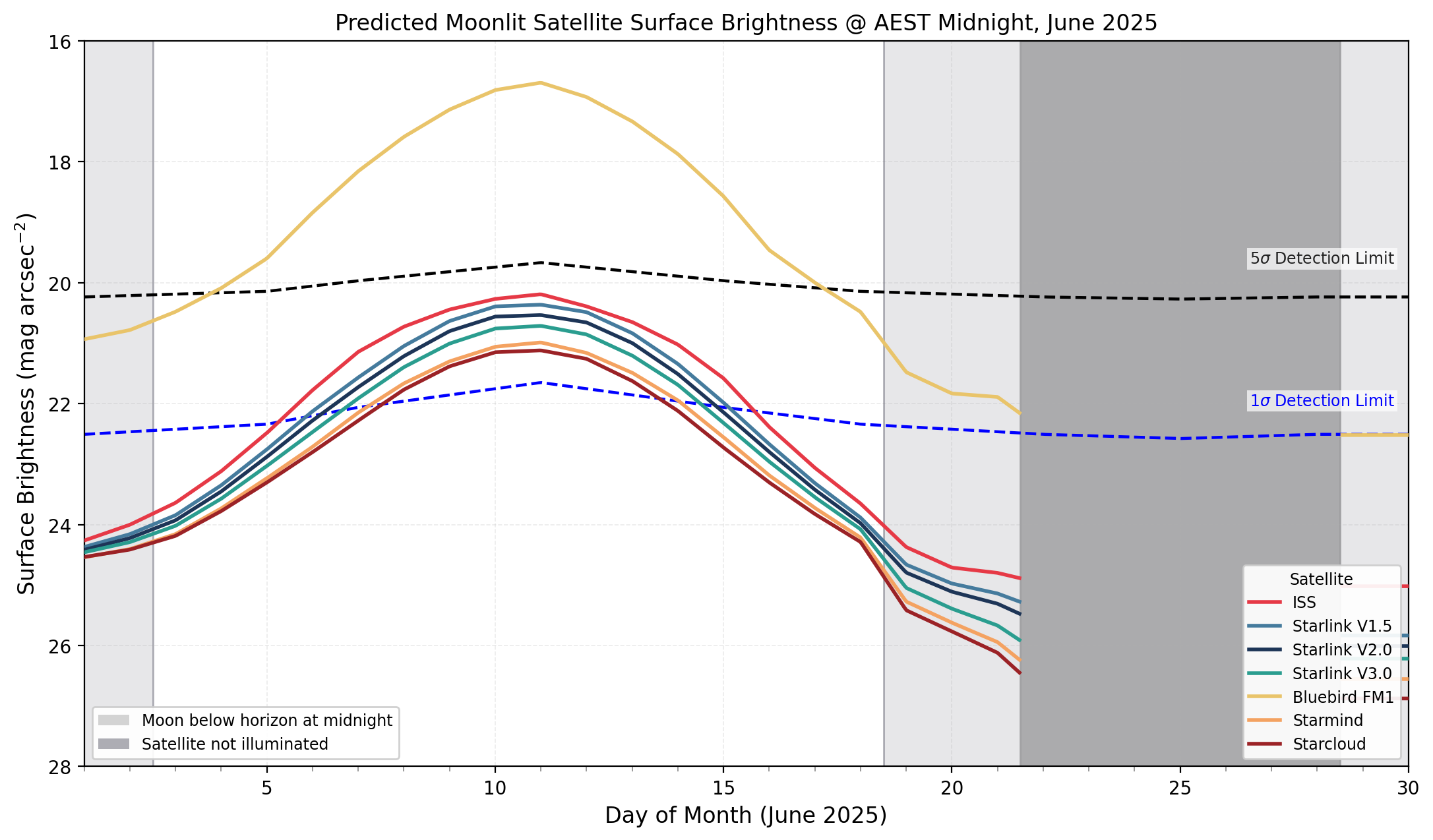}
    \caption{The modelled surface brightness of the satellites considered in this work each day for a representative month of June 2025. Surface brightness estimates are taken from an averaged 1 hour period either side of midnight, from sampling points across the whole sky at the time of observation. Shaded zones show where the Moon is below the horizon at midnight but the satellite is still illuminated, and when the satellite is not illuminated. $1\sigma$ and $5\sigma$ surface brightness limited are shown as a function of day of the month (and sky brightness from the Moon) for the telescope system considered in this work.}
    \label{fig:all_phase_moon}
\end{figure}

While these results are promising for full Moon conditions, we now explore if Moonlight can be used to meaningfully extended SDA observations of satellites at night across an entire month. To do so we considered the surface brightness of the satellites considered in this work as a function of lunar phase angle. For large resolved satellites like the ISS for which the angle subtended on the sky is larger than the PSF, important factors that determine how easily that object will be detected when illuminated by Moonlight are the scattering surface properties of the materials of the satellite, the orientation of the satellite, and the brightness of the source of illumination. As demonstrated in \autoref{surface_bright}, the surface brightness of the satellite becomes independent of the distance to the source, and the surface area. In \autoref{fig:all_phase_moon} we use {\texttt{lumos-sat}} to explore the surface brightness of the satellites considered in this work as a function of solar phase angle throughout the representative month of June 2025. The surface brightness of each satellite is calculated on a per day basis, by averaging over a 1 hour period at local midnight of each 24 hour period throughout the month by taking the median of the surface brightness calculated at uniformly distributed points across the entire sky to create a representative median surface brightness for that night. The brightness of Moon as a function of phase angle is calculated using \autoref{moon_bright}. Approximately half of all nights in the month do no have the Moon above the horizon at midnight, satellites may still be illuminated by it, similarity to twilight illuminated conditions in traditional SDA observing 1-2 hours after sunset and shown in \autoref{fig:all_day}. For $\sim6$ of the nights of the month, the satellites are not illuminated at all by the Moon at midnight in their respective orbital altitudes. 
\noindent
\\
\\
The $1\sigma$ and $5\sigma$ detection limits for a representative tracked 10s exposure are calculated for the 24" Planewave telescope used in this work as a function of sky brightness using the exposure time calculator framework {\texttt{Gunagala}} \citep{robitaille_astrohuntsmangunagala_2022}. Representative sky brightnesses are taken from the DECam exposure time calculator\footnote{\url{https://noirlab.edu/science/programs/ctio/instruments/Dark-Energy-Camera/User-Guide/Exposure-Time-Calculator-ETC-0}}. The $1\sigma$ and $5\sigma$ surface brightness limits for the telescope used in this work are found to be $21.65 mag/arcsec^2$ and $19.67 mag/arcsec^2$ respectively at full Moon conditions, and $22.56 mag/arcsec^2$ and $20.27 mag/arcsec^2$ at new Moon conditions (for a dark site). The result of considering the BRDF's of the Starmind, Starcloud and ISS satellites chassis and solar panel components to be the same as what has been measured on sky for Starlink, all of these satellites exhibit a similar peak surface brightness, modified by the relative size of the bright chassis component with respect to the solar panels and all have an average surface brightness between $22mag/arcsec^2$ and $20 mag/arcsec^2$. The ISS is found to have a peak surface brightness of $20.05 mag/arcsec^2$ at middnight during full Moon, which is within the measured peak surface brightness of the ISS observed in this work of $19.65\pm1.00 mag/arcsec^2$. All of these satellites are above the $1\sigma$ detection threshold for our telescope system for a maximum of 11/30 days of the month at midnight.
\\
\\
Earendil-1 is also modelled, but because of it's highly spectacular surface and orbital geometry, the narrow beam (5.8km diameter pointing directly nadir) does not intersect with the modelled observer when considering a Sun tracking concept of operations, and so does not feature on this plot. Interestingly, Bluebird FM1 has the highest peak surface brightness of $16.34mag/arcsec^2$, likely due to its high albedo, modelled lambertian reflective surface (as opposed to the Starlink BRDF), and orbital geometry where the satellites large surface area is always facing nadir, maximising Lunar-Earthshine reflectance. It should be noted these results come with a caveat and require verification within orbital brightness measurements to confirm that a lambertian scattering surface is a correct approximation for the Bluebird FM1 satellite. Bluebird FM1 is above the detection threshold for all nights where the satellite is illuminated by the Moon at midnight. These results indicate that even for a modest telescope system such as the one used in this work, if an object like the Bluebird FM1 can be tracked for 10s, then Moonlight and Lunar-Earthshine can be used to monitor the object even at midnight when traditional optical SDA operators would not be attempting to track such objects. These findings are echoed in works such as \cite{nandakumar_high_2023} and \cite{cole_initial_2025} which raise concern from astronomers over the high optical brightness of large BlueWalker and Bluebird generation of satellites due to their large nadir facing surface area and white, high albedo coating.

\subsection{Implications for Astronomy}

The prospect of a satellite being illuminated by a light source throughout an entire night is of concern to Astronomers who rely on the Earth's shadow to prevent illumination of LEO objects during the most productive hours of the night for astronomical research. Unlike tracked observations of satellites, the effective surface brightness of streaks going through an astronomy image is strongly dependant on the surface area of the satellite in question, and as a result, the advent of larger satellites such as orbital data centres, may be of concern. 
\begin{figure}[t]
    \centering
    \includegraphics[width=0.9\linewidth]{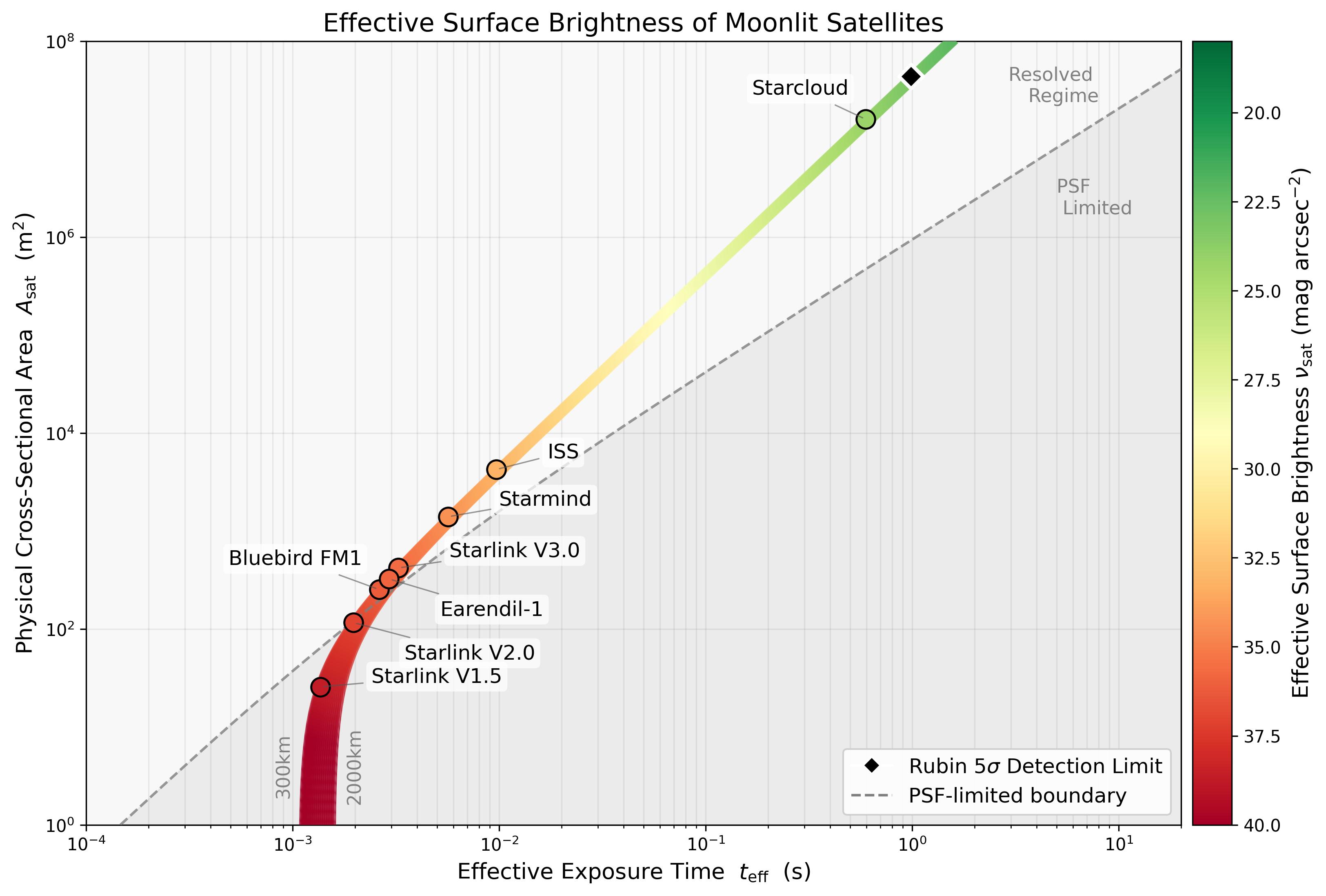}
    \caption{The effective surface brightness of a satellite trail in a 15s Rubin exposure as a function of satellite cross-sectional area and effective exposure time. The calculated surface brightness limit for a single Rubin 15s exposure at full Moon is shown as the black diamond data point. The width of the plotted line is the spread in effective exposure time for orbital altitudes from 300-2000km.}
    \label{fig:teff_area}
\end{figure}
\noindent
The property which dictates the SNR of a satellite streak in an image is the effective exposure time, or the total time for which each pixel in the detector is illuminated by the satellite source. This is governed by the surface area of these satellites with respect to the observer, and the orbital altitude. The physical cross-sectional area $A_{\rm sat}$ of a satellite and its effective exposure time $t_{\rm eff}$ are related at orbital altitude $d_{sat}$ by:

\begin{equation}
    t_{\rm eff} = 2\sqrt{\frac{A_{\rm sat}}{\pi} \cdot 
    \frac{R_\oplus + d_{\rm sat}}{GM_\oplus}}
    \label{eq:teff_simple}
\end{equation}
\noindent
Where G is the gravitational constant, and $M_\oplus$ is the mass of the Earth. As $R_\oplus = 6371$\,km $> d_{\rm sat}$ across the range of typical LEO orbits, the altitude term varies by only $\sim$12\% between 300--2000\,km, whilst $t_{\rm eff}$ scales directly with $\sqrt{A_{\rm sat}}$. A satellite with 100$\times$ the cross-sectional area therefore produces a $10\times$ longer 
effective exposure time and a streak 100$\times$ brighter in surface brightness, regardless of its orbital altitude. This is illustrated in \autoref{fig:teff_area}, where the narrow width of the shaded band across the  300--2000\,km LEO range is negligible compared to the orders-of-magnitude variation in surface brightness driven by the satellites surface area alone.
\\
\\
To explore how this may impact astronomy images for the Vera Rubin Observatory (Rubin herein; \citealt{team_vera_2026}), the effective satellite trail surface brightness is plotted as a function of cross-sectional area of the satellite, and the effective exposure time of satellite trail in \autoref{fig:teff_area}. The quadratic ($A_{\rm sat} \propto t_{\rm eff}^2$) relationship is shown in the slope of the log-log plot, with the width of line being the span in the relationship between 300-2000km orbital altitudes. We take the diameter of the telescope aperture to match Rubin at 8.4m, and the typical seeing to be 0.7 arcseconds \citep{crenshaw_using_2024}.
\\
\\
Each satellite considered in this work is plotted on the line to illustrate the effective surface brightness of each in a typical 15s exposure of the sky in Rubin g' band-pass. To calculate the limiting surface brightness for Rubin in a 15s exposure, we use the Python implementation of Rubin-sim \footnote{\url{https://github.com/lsst/rubin_sim}} with a night sky brightness equal to full Moon estimate of 19.4 mag from the DECam exposure time calculator for DECam g' band-pass observations (very close to the Rubin g' band-pass) and a seeing value of 0.7 arcseconds. This is found to be $23.13 \, mag/arcsec^2$ for an object in a typical 15s exposure for $5\sigma$ observations ($23.16 \,mag/arcsec^2$ for Rubin r' and $23.12 \,mag/arcsec^2$ for Rubin i').
\\
\\
As an example, considering our median ISS total integrated brightness estimate of $12.02\pm1.00$ V band magnitudes from on-sky measurements, an average orbital altitude of 400km, and a total surface area $4250.25m^2$ (object at zenith, solar panels facing nadir so this is a reasonable approximation), the calculated surface brightness is $19.70\pm 1.00 \pm mag/arcsec^2$, and the effective surface brightness is $35.78\pm 1.00 \pm mag/arsec^2$. This is well below the Rubin surface brightness limit, and will not be detected. 
\\
\\
We find that for even the largest satellites like Starcloud, thankfully are likely to be too faint to significantly disrupt astronomy observations when lit by the Moon in full Moon conditions.  

\section{Conclusions}
In this work we have demonstrated observations of satellites illuminated exclusively by the light of the Moon, and Lunar-Earthshine reflecting off the surface of the Earth at night for the first time. Critically, this finding opens the opportunity to increasing the productive of optical ground based SDA observations during full Moon conditions to utilising the full 24 hour period for some satellites, even with a modest 0.6m telescope used in this work. We find that the mean brightness of the ISS at night is $12.02\pm1.00$ V-mag, and use these observations to validate our extended version of satellite brightness modelling software {\texttt{lumos-sat}} which shows agreement with our observations. We note that from these results, the ISS is found to be $13.12 \pm 1.30$ magnitudes fainter at night than when observed during the day, which is within the order of magnitude estimate of 14 magnitudes from first principle comparisons of the difference in brightness between the Sun and the Moon.
\\
\\
We also use the brightness modelling software Ansys STK to explore the importance of considering Lunar-Earthshine in night time observations of satellites, which at times can be the only illumination source for large nadir facing surfaces of a satellite (like the solar panels at local midnight) and can increase the brightness of a satellite by up to $240\times$ compared to starlight alone, for highly reflective components of the ISS like the radiators. We predict that observations of satellites like the ISS may be possible for up to 11/30 days of the month with our 0.6m telescope, improving the duty cycle of traditional optical ground based SDA observations. Finally, we find that is unlikely that moonlit satellites will cause concern for astronomical observatories like Rubin, even for the largest planned orbital data centre satellites like Starcloud and Starmind. 
\\
\\
Future work will include expanding the coverage of our observations across the full 24 hour period during full Moon conditions. We will also expanded the sampling of lunar phase angles by conducting observations over multiple day of the month. We also plan to improve the photometry of the system by installing photometric filters, and improve dome tracking to prevent any possible vignetting. In addition, we plan to extend {\texttt{lumos-sat}} to include more detailed multi-component models (in particular incorporating surface brightness natively and adapt surface area for differences in line of sight distance and viewing angles), more accurate consideration of atmospheric extinction, and parallelise computations to improve the efficiency and reduce computation time for more complex detailed simulations.
\\
\\
This work has further extended the findings of \cite{caddy_daytime_2025}, by exploring another challenging optical observational regime that traditionally was not considered to be productive for optical SDA operators. Like daytime observations, Moonlit observations at night present challenging bright sky backgrounds compared to the target brightness, but with appropriate observing techniques and a comprehensive understanding of illumination conditions demonstrated in this work, these challenges can be overcome. As a result of this work, it is clear that Moonlight and Lunar-Earthshine are a non-negligible illumination source for satellites, and may be exploited to increase the productivity of traditional ground based optical SDA facilities.

\backmatter
\bmhead{Acknowledgements}

The authors would like to thank the manager of the Macquarie University Observatory, Mr. Adam Joyce, for his continued support of space domain awareness projects conducted by the team. 
\\
\\
In addition the authors would like to acknowledge the traditional owners of the land on which the Macquarie University Observatory is located, the Wallumattagal Clan of the Dharug Nation - whose cultures and customs have nurtured, and continue to nurture, this land since time immemorial.

\section*{Declarations}

\begin{itemize}
\item \textbf{Funding}: This research is supported in part by the Australian Office of National Intelligence through the National Intelligence and Security Discovery Research Grant award NI230100162. 
\item \textbf{Competing Interests}: The authors declare there are no competing interests relating to this work. 
\end{itemize}

\newpage 

\begin{appendices}

\section{ISS Component Dimensions}\label{secA1}

\begin{table}[h!!!]
\small
\begin{tabular}{lccc}
\toprule
\rowcolor{gray!20}
Component & Length (m) & Diameter (m) & Number \\
\midrule
Destiny & 9.2 & 4.3 & 1 \\
Columbus & 6.9 & 4.5 & 2 \\
Hope (PM) & 11.2 & 4.4 & 3 \\
Hope (ELM-PS) & 3.9 & 4.4 & 4 \\
Hope (EF) & 5.6 & 5 & 5 \\
Node 1 Unity & 5.5 & 4.3 & 6 \\
Node 2 Harmony & 6.7 & 4.3 & 7 \\
Node 3 Tranquility & 6.7 & 4.3 & 8 \\
Airlock & 5.5 & 4 & 9 \\
Cupola* & 3 & 1.5 & 10 \\
PMM & 6.67 & 4.5 & 11 \\
Zarya & 12.99 & 4.1 & 12 \\
Zarya solar panels & - & (28 m$^2$ area) & 13 \\
Docking compartment & 4.9 & 2.55 & 14 \\
MRM2 & 4.9 & 2.55 & 15 \\
MRM1 & 6.0 & 2.35 & 16 \\
Zvezda & 13.1 & 4.2 & 17 \\
Zvezda solar panels & 29.7 & $\sim 4$ & 18 \\
PMAS 1 & 1.86 & 1.63 & 19 \\
PMAS 2 & 1.86 & 1.63 & 20 \\
PMAS 3 & 1.86 & 1.63 & 21 \\
\bottomrule
\caption{Main modules of the ISS. Components labeled with * are not considered as they to not add to the total scattering surface area when viewed from Earth.}
\end{tabular}
\end{table}

\begin{table}[h!!!]
\small
\begin{tabular}{lccc}
\toprule
\rowcolor{gray!20}
Component & Length (m) & Diameter (m) & Number \\
\midrule
Truss & 108.5 & 4.3 & 22 \\
Main solar panels & -- & (2500 m$^2$ area) & 23 \\
Radiators (solar) x4 & 13.6 & 3.12 & 24 \\
Radiator (main) x6 & 23.3 & 3.4 & 25 \\
\bottomrule
\caption{Components associated with the main frame or truss of the ISS}
\end{tabular}
\end{table}

\begin{table}[h!!!]
\small
\begin{tabular}{lccc}
\toprule
\rowcolor{gray!20}
Component & Length (m) & Diameter (m) & Number \\
\midrule
Progress 91 Solar Array & 10.7 & 1.2 & 26 \\
Progress 91 & 7.4 & 2.7 & 27 \\
Soyuz MS 27 Solar Array & 10.6 & $\sim1.2$ & 28 \\
Soyuz MS 27 & 7 & 2.7 & 29 \\
Crew 10 Dragon & 5.1 & 3.66 & 30 \\
\bottomrule
\caption{Visiting vehicles at the ISS correct for June 2025 when the space station was observed in this work.}
\end{tabular}
\end{table}

\end{appendices}

\bibliography{sn-bibliography}

\end{document}